\documentclass[preprint,12pt]{elsarticle}
\usepackage{xcolor}

\usepackage{amssymb}
\usepackage{amsmath}

\usepackage{acronym}
\usepackage{orcidlink}

\journal{Elsevier Internet of Things}

\begin{document}

\acrodef{ae}[AE]{Autoencoder}
\acrodef{ai}[AI]{Artificial Intelligence}
\acrodef{ann}[ANN]{Artificial Neural Network}
\acrodef{bilstm}[Bi-LSTM]{Bidirectional Long Short-Term Memory}
\acrodef{ble}[BLE]{Bluetooth Low Energy}
\acrodef{cnn}[CNN]{Convolutional Neural Network}
\acrodef{cpu}[CPU]{Central Processing Unit}
\acrodef{dl}[DL]{Deep Learning}
\acrodef{dt}[DT]{Decision Tree}
\acrodef{dnn}[DNN]{Deep Neural Network}
\acrodef{drl}[DRL]{Deep Reinforcement Learning}
\acrodef{elm}[ELM]{Extreme Learning Machine}
\acrodef{et}[ET]{Extra Trees}
\acrodef{gb}[GB]{Gradient Boosting}
\acrodef{gpu}[GPU]{Graphics Processing Unit}
\acrodef{hb}[HB]{Histogram-based Gradient Boosting}
\acrodef{iiot}[IIoT]{Industrial Internet of Things}
\acrodef{iomt}[IoMT]{Internet of Medical Things}
\acrodef{iot}[IoT]{Internet of Things}
\acrodef{lcf}[LCF]{Linear-Curve Fitting}
\acrodef{lr}[LR]{Linear Regression}
\acrodef{lstm}[LSTM]{Long Short-Term Memory}
\acrodef{ml}[ML]{Machine Learning}
\acrodef{mlp}[MLP]{Multilayer Perceptron}
\acrodef{qcf}[QCF]{Quadratic-Curve Fitting}
\acrodef{qos}[QoS]{Quality of Service}
\acrodef{rf}[RF]{Random Forest}
\acrodef{rl}[RL]{Reinforcement Learning}
\acrodef{svm}[SVM]{Support Vector Machine}

\begin{frontmatter}



\title{Machine Learning on the Edge for Sustainable IoT Networks: A Systematic Literature Review\tnoteref{funding}} 
\tnotetext[funding]{The present work has received funding from the European Union's Horizon 2020 Marie Skłodowska Curie Innovative Training Network Greenedge (GA. No. 953775).}


\author[1]{Luisa Schuhmacher\orcidlink{0000-0003-3293-7933}}
\author[1]{Jimmy Fernandez Landivar\orcidlink{0000-0002-4904-5256}}
\author[1]{Ihsane Gryech\orcidlink{0000-0001-5288-4205}}
\author[1]{Hazem Sallouha\orcidlink{0000-0002-1288-1023}}
\author[2]{Michele Rossi\orcidlink{0000-0003-1121-324X}}
\author[1,3]{Sofie Pollin\orcidlink{0000-0002-1470-2076}} 

\affiliation[1]{organization={WaveCoRE, Department of Electrical Engineering (ESAT), KU Leuven},
            addressline={Kasteelpark Arenberg 10 postbus 2440}, 
            city={Leuven},
            postcode={3001}, 
            country={Belgium}}
\affiliation[2]{organization={Department of Information Engineering, University of Padova},
            addressline={Via Gradenigo 6/b}, 
            city={Padua},
            postcode={35131}, 
            country={Italy}}
\affiliation[3]{organization={IMEC},
            addressline={Kapeldreef 75}, 
            city={Leuven},
            postcode={3001}, 
            country={Belgium}}

\begin{abstract}
The Internet of Things (IoT) has become integral to modern technology, enhancing daily life and industrial processes through seamless connectivity. However, the rapid expansion of IoT systems presents significant sustainability challenges, such as high energy consumption and inefficient resource management. Addressing these issues is critical for the long-term viability of IoT networks. Machine learning (ML), with its proven success across various domains, offers promising solutions for optimizing IoT operations. ML algorithms can learn directly from raw data, uncovering hidden patterns and optimizing processes in dynamic environments. Executing ML at the edge of IoT networks can further enhance sustainability by reducing bandwidth usage, enabling real-time decision-making, and improving data privacy. Additionally, testing ML models on actual hardware is essential to ensure satisfactory performance under real-world conditions, as it captures the complexities and constraints of real-world IoT deployments. Combining ML at the edge and actual hardware testing, therefore, increases the reliability of ML models to effectively improve the sustainability of IoT systems. The present systematic literature review explores how ML can be utilized to enhance the sustainability of IoT networks, examining current methodologies, benefits, challenges, and future opportunities. Through our analysis, we aim to provide insights that will drive future innovations in making IoT networks more sustainable.
\end{abstract}









\begin{keyword}
Internet of Things \sep Machine Learning \sep Sustainability \sep Energy Efficiency \sep Edge Computing \sep Hardware \sep Testbeds \sep Systematic Literature Review


\end{keyword}

\end{frontmatter}



\section{Introduction}
\label{introduction}

The \ac{iot} has become a cornerstone of modern technology, permeating various aspects of daily life and industrial processes. By connecting a myriad of devices, from household appliances to industrial machinery, \ac{iot} networks facilitate real-time data exchange and automation, leading to increased efficiency and convenience \cite{prabhu_iot_2024}. However, the exponential growth of \ac{iot} systems presents significant sustainability challenges. The vast number of connected \ac{iot} devices, in sum, contributes to substantial energy consumption, which in turn affects carbon emissions and environmental sustainability \cite{prakash_green_2023}. Furthermore, managing resources such as bandwidth and computational power in \ac{iot} networks often leads to inefficiencies and increased operational costs \cite{raghavendar_robust_2023}. Addressing these sustainability and operational cost issues is paramount to ensuring the environmental and economic viability of \ac{iot} networks in the long term. To this end, optimization on the software side regarding the allocation of computational resources and \ac{iot} network operation becomes a crucial area to investigate.

\ac{ml}, a subset of Artificial Intelligence, has demonstrated remarkable success across various domains, including healthcare, finance, and transportation \cite{mukhamediev2022review}. By utilizing sophisticated algorithms and vast amounts of data, \ac{ml} can uncover patterns, make predictions, and optimize processes in ways that were previously unimaginable. In healthcare, for instance, \ac{ml} algorithms have improved diagnostic accuracy and personalized treatment plans \cite{venigandla2022integrating}. In finance, \ac{ml} models enhance fraud detection and risk management \cite{tatineni2024enhancing}. Similarly, in transportation, \ac{ml} has enabled the development of autonomous vehicles and optimized logistics \cite{chung2021applications, khayyam2020artificial}. The proven capabilities of \ac{ml} in these fields suggest its potential to address complex challenges, such as those found in \ac{iot} networks.

Compared to traditional statistical methods, \ac{ml} offers several distinct advantages that are particularly beneficial for \ac{iot} networks. While statistical methods rely on predefined models and assumptions about data distribution, \ac{ml} algorithms can learn directly from raw data. This makes them more flexible and adaptive to complex and dynamic environments \cite{averbuch2022applications}. This adaptability is crucial for \ac{iot} systems, which generate diverse and often intermittent and non-stationary data. Additionally, \ac{ml} techniques can uncover hidden patterns and relationships within the data that statistical methods might miss. More innovative solutions for optimizing \ac{iot} operations can be achieved through this \cite{alkhayyal2024recent}. However, \ac{ml} usually requires a lot of data which implies collecting and transferring vast amounts of data over the network \cite{zhou2017machine}.

One promising development, therefore, is the execution of \ac{ml} on the edge of \ac{iot} networks. Edge computing involves processing data near the source of data generation rather than in centralized cloud servers \cite{yu2017survey}. This approach offers several benefits for sustainability \cite{ning2018green}. Firstly, edge computing significantly lowers bandwidth usage and associated energy consumption by reducing the need to transmit large data volumes to the cloud. Secondly, edge computing enhances the responsiveness and efficiency of \ac{iot} systems by enabling real-time data processing and decision-making. This is particularly important for applications requiring immediate actions, such as predictive maintenance and energy management. Moreover, processing data locally at the edge can enhance data privacy and security, further bolstering the overall robustness of \ac{iot} networks.

The importance of testing machine learning models on hardware at the edge cannot be stressed enough. While useful for initial model development, simulated environments often fail to capture the complexities and constraints of real-world \ac{iot} deployments \cite{almutairi2024advancements}. Actual hardware testing ensures that \ac{ml} models perform satisfactorily under the specific conditions they will encounter, such as limited computational resources, power constraints, and varying environmental conditions. It also allows for the fine-tuning of models to achieve the best trade-off between accuracy and resource usage. Moreover, testing on actual hardware helps identify potential issues related to hardware-software integration, which is crucial for the seamless operation of \ac{iot} networks. This hands-on approach is also essential for validating the effectiveness of \ac{ml} models in improving the sustainability of \ac{iot} systems.

Linking \ac{ml} with \ac{iot} networks presents a promising avenue for enhancing sustainability. Specifically, \ac{ml} can be employed to optimize energy usage by predicting and adjusting the power consumption of \ac{iot} devices, thereby reducing overall energy demand \cite{tekin2023energy, schuhmacher2023ecoble}. Resource allocation can be improved through \ac{ml} algorithms that dynamically manage network resources based on real-time data, minimizing waste and maximizing efficiency \cite{djigal2022machine}. Additionally, \ac{ml} can aid in predictive maintenance, identifying potential failures in \ac{iot} systems before they occur, thus reducing downtime and the need for resource-intensive repairs \cite{sami2023forecasting}. By integrating \ac{ml} into \ac{iot} networks, especially through edge computing and rigorous testing on actual hardware, we can not only enhance their operational efficiency but also significantly mitigate their environmental impact.

In this context, a few related surveys exist. Some provide a broader perspective on deploying machine learning in resource-constrained IoT environments through edge computing to mitigate network congestion, latency, and privacy concerns \cite{merenda, murshed}. However, these are conventional surveys rather than systematic literature reviews (SLRs). Others are more specialized and closely aligned with our work, even adopting a systematic approach, such as \cite{survey1}. In this work, the authors present a mapping of \ac{ai}-based solutions to achieve energy sustainability together with better \ac{qos} in the different layers of \ac{iot} networks. Although this paper is significantly related to our field of interest, it distinguishes itself in three major ways from our systematic review: Firstly, next to energy efficiency, selected papers further needed to improve the \ac{qos}. Secondly, the authors considered \ac{ai} broadly, which includes \ac{ml} but also other \ac{ai} methods such as Swarm Intelligence. Lastly, they focused on optimizing \ac{iot} networks and on the methodology to pursue it while we further investigate actual hardware implementations and the specifications of the therefore established \ac{iot} networks. Trends and methodologies for energy management in \ac{iot} networks were investigated in \cite{slr}. Though this is related, the authors focus on energy management, while we broaden it to sustainability in general. Moreover, they did not specifically look into \ac{ml} methods, which is a key research focus of this paper. Further, a comparative study of existing algorithms to enhance the energy efficiency of \ac{iot} networks was conducted in \cite{comp_study}. Unlike our systematic literature review, they also include non-\ac{ml} methods and investigate simulators instead of actual hardware implementations. 


While the related surveys and reviews mentioned provide relevant information, they do not specifically address the problem of evaluating simulation-based conclusions on for instance energy consumption with hardware-based results.  Simulation models can be very intricate, some modelling the real world in much more detail than others; however, most of them are based on simplifying assumptions. In general, it is impossible to define a complete model. To advocate performance verification on actual hardware and show different ways of doing it, we present a systematically collected range of research evaluated on actual hardware. This research can start small by presenting a prototype, contain a network implementation in a testbed environment, or utilize a full-fledged implementation in an \ac{iot} network. To be specific, this systematic literature review is centred on the intersection of machine learning and \ac{iot} sustainability evaluated on real hardware on the edge. Thus, we offer a comprehensive analysis of current research, methodologies, and applications. We will explore how \ac{ml} algorithms are being utilized to create more sustainable \ac{iot} networks, identify the benefits and challenges associated with these approaches, and highlight emerging trends and future opportunities. Through this analysis, we seek to provide valuable insights and a foundational understanding that will drive future innovations in making \ac{iot} networks more sustainable. Specifically, the main contributions of this paper are as follows:
\begin{itemize}
    \item Conduction of a systematic review using the PRISMA methodology to ensure a structured, transparent, and replicable process.
    \item Adherence to a predetermined protocol to maintain objectivity and impartiality throughout the review process.
     \item Extraction and summary of papers from the literature that proposes \ac{ml} models to improve the sustainability of \ac{iot} networks and validate their proposed methods by actual hardware implementation.
    \item Elucidation of different aspects of the network to improve sustainability in, using \ac{ml}.
    \item Specification of the \ac{iot} networks regarding the hardware used and application implemented to validate sustainability improvements.
    \item Investigation of which \ac{ml} methods have been used and compared with each other, and which showed the best performance for a given setup.
    \item List metrics used to evaluate the improvement of the network's sustainability.
    \item Identification of gaps in the literature and suggestions for future work to advance the sustainability of \ac{iot} networks.
\end{itemize}

The remainder of the paper is organized as follows. Section \ref{methods} covers the materials and methods, including the research questions, data sources, and inclusion criteria. Section \ref{results} provides a summary of the review findings and a detailed comparison of the methods used in the surveyed literature. A discussion about the identified gaps, directions for future research, and practical implications is provided in Section \ref{discussion}. Finally, Section \ref{conclusion} presents the conclusions.

\section{Materials and methods}\label{methods}

\begin{figure}[!htb]
    \centering
    \includegraphics[width=0.7\textwidth]{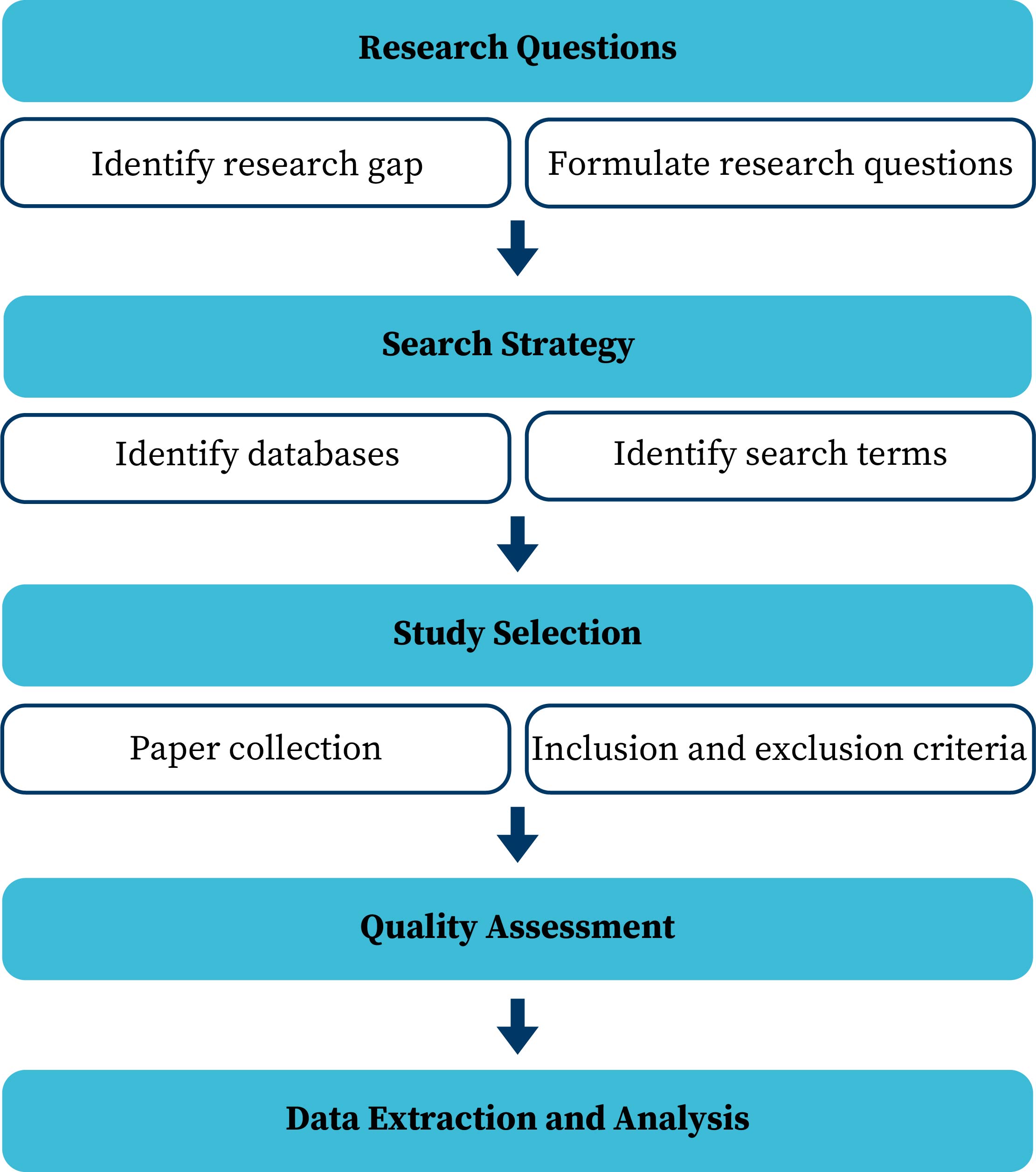}
    \caption{PRISMA Methodology followed in the systematic review process \cite{SLRig, SLRuv}}
    \label{fig:methodology}
\end{figure}

This work presents a Systematic Literature Review (SLR), employing the PRISMA (Preferred Reporting Items for Systematic Reviews and Meta-Analyses) methodology (Fig. \ref{fig:methodology}) \cite{page2021prisma}, and thus ensuring a structured, transparent, and replicable process. The study enhances rigour and reproducibility by analyzing indexed research from SCOPUS over the past decade (2013 – 2023), facilitating validation by other researchers. Explicit inclusion and exclusion criteria are established to minimize selection bias, guaranteeing comprehensive consideration of pertinent research. Adherence to a predetermined protocol maintains objectivity and impartiality throughout the review process, distinguishing this systematic review from traditional approaches.   

The SLR procedure encompasses three primary phases: (i) Identification, (ii) Screening, and (iii) Eligibility. Upon formulating the research questions, the Identification phase establishes the search strategy, specifying the selection of data sources and extraction methods for gathering pertinent papers. In the screening and eligibility phases, inclusion and exclusion criteria, aligned with the review's specific requirements and scope, are delineated. During these phases, papers are filtered based on titles and abstracts (screening) and full-text (eligibility). Fig. \ref{fig:flowchart} provides a comprehensive overview of the outcomes at each stage of the PRISMA process. Subsequently, responses to the research questions are synthesized, and the challenges, opportunities, and limitations are underscored. These processes are elaborated on in the following sections.

\subsection{Research Questions}

\ac{ml} plays a crucial role in enhancing energy efficiency within the \ac{iot} framework. The deployment of \ac{ml} algorithms at the edge of \ac{iot} networks can bring forth numerous energy-saving strategies \cite{ning2018green, li2022energy, chen2021energy}. \ac{ml} can empower real-time decision-making at the edge and diminish the constant necessity for data transmission to centralized servers. This localized processing capacity allows for direct filtering, compression, and analysis of data on \ac{iot} devices. It thereby reduces the quantity of information transmitted over the network and subsequently curbs energy consumption \cite{bebortta2021robust, cai2019iot}. \ac{ml}'s contribution could extend to predictive maintenance, enabling early detection of equipment failures or anomalies, thus averting unnecessary energy usage and downtime. Additionally, adaptive power management and personalized energy profiles, guided by \ac{ml} models, can guarantee that \ac{iot} devices operate in the most energy-efficient states based on usage patterns and individual preferences. In conclusion, integrating \ac{ml} at the edge of \ac{iot} networks has the potential to be a vital tool in advancing sustainable practices. This integration can optimize resource allocation, reduce data transfer, and promote intelligent, energy-conscious decision-making at the edge. However, there is no strict path to enhancing sustainability, leaving the option of green-washing research. Due to the hype about sustainability, it occurs that authors state that their conducted research improves sustainability while no credible metric is being used for assessment, or not all components of the method are being examined \cite{gatti2019grey}.

To comprehend the current landscape in this context and to avoid including potentially green-washed research, our systematic review aims to identify critical research questions and seek relevant answers via a thorough investigation. The three research questions formulated for this systematic review are:

\begin{itemize}
    \item \textbf{RQ1:} How can \ac{ml} contribute to enhancing energy efficiency and sustainability at the edge of \ac{iot} networks?
    \item \textbf{RQ2:} What \ac{iot} network specifications were used to test the proposed \ac{ml} methods?
    \item \textbf{RQ3:} Which \ac{ml} tools are employed at the edge to enhance the energy efficiency and sustainability of \ac{iot} networks?
\end{itemize}

These inquiries have been formulated to fulfill the primary objective of this paper: conducting a systematic literature review on the application of machine learning methods in enhancing energy efficiency and sustainability within the \ac{iot}. Additionally, this paper seeks to offer a comprehensive synthesis of the \ac{ml}-driven contributions to the sustainability of \ac{iot} devices, including the tools and models employed for this purpose. Furthermore, the paper provides insights into the current state of research in the field, highlighting key challenges and potential practical implications.

\subsection{Search Strategy}

To investigate the research questions, we leveraged the SCOPUS dataset \cite{scopus}, as it is recognized as the world's largest abstracting and indexing data\-base, known for its continuous daily updates \cite{Schotten}. The search for relevant publications was initiated on August 8th, 2023.

Keywords and queries were formulated according to the requirements of the Scopus scientific database using the research questions. The following research query was employed: 
\textbf{ALL ( "IOT" ) OR ALL ( "Internet of Things" ) AND ALL ( "energy efficiency" ) OR ALL ( "sustainability" ) AND ALL ( "machine learning" ) OR ALL ( "deep learning" ) AND ALL ("edge computing" ) AND PUBYEAR \textgreater 2012 AND PUBYEAR \textless 2024 AND ( LIMIT-TO ( LANGUAGE , "English" ) ) AND ( LIMIT-TO ( DOCTYPE , "ar" ) OR LIMIT-TO 
( DOCTYPE , "cp" ) )}.

The search time frame covered a decade of publications, spanning January 2013 to August 2023. The initial search found 5,827 documents. These documents were added to Rayyan, a smart research collaboration platform designed to streamline the process of literature reviews and systematic reviews \cite{Rayyan}. On this platform, 40 duplicates were identified, and only 7 were verified as true duplicates before being removed.
The remaining 5820 studies were further processed according to the inclusion and exclusion criteria (Fig. \ref{fig:flowchart}).

\subsection{Study Selection and Data Extraction}
\begin{figure}[h]
    \centering
    \includegraphics[width=0.9\textwidth]{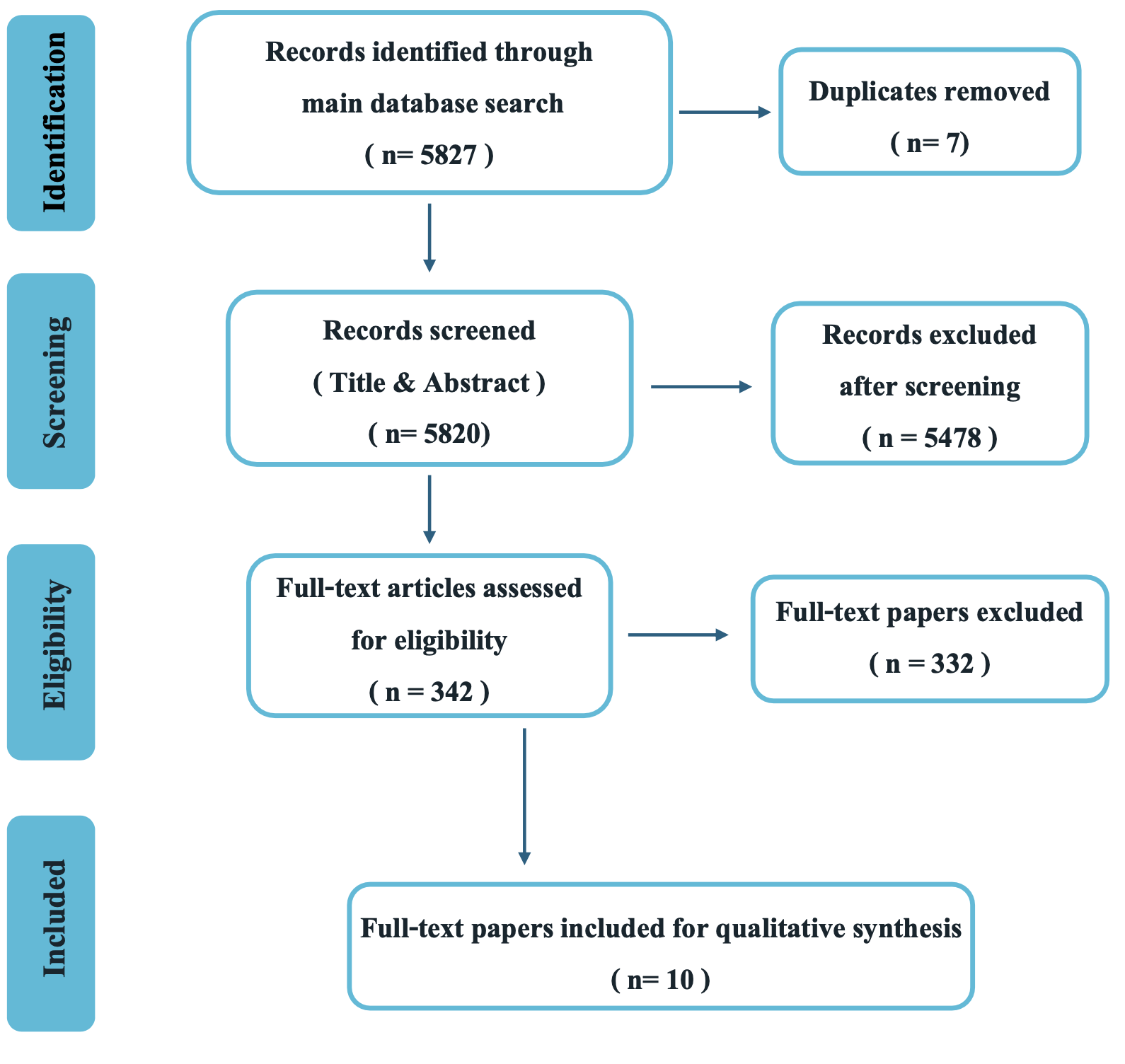}
    \caption{Flow Diagram for the selection of the literature reviewed. 
    }
    \label{fig:flowchart}
\end{figure}
\begin{table}[]
\begin{tabular}{|l|l|}
\hline
\textbf{Inclusion Criteria} & \textbf{Exclusion Criteria} \\ 
\hline
- Within scope of research & - Does not include edge computing. \\
questions. & - Proposed method does not positively \\
& impact energy efficiency or sustainability \\
& of \ac{iot} networks, or is not discussed. \\
& - Only \ac{ml}-specific hardware optimization. \\
& - Makes application on the edge more \\
& energy-efficient, but not the network itself. \\ 
\hline
- Application of \ac{ml} on the & - No direct application on the edge, solely \\
edge. & research on energy-efficient \ac{ml} methods \\
& for edge computing (e.g., optimized \\
& federated learning algorithm). \\
\hline
- Actively improves energy & - Only meets resource constraints (e.g., \\
efficiency or sustainability. & computational constraints by specific \\
& hardware, or financial constraints). \\
\hline
- Explicit usage of an \ac{iot} & - Usage of technologies that can be used \\
network. & for \ac{iot} networks but not embedded in \\
& an \ac{iot} network (e.g., UAVs, 5G, Mobile \\
& Edge Computing...). \\
\hline
- Evaluation of \ac{ml} on an & - Only simulation-based evaluation, no \\
edge device. & real hardware implementation. \\
\hline
- Within the time frame of & - Outside the time frame of publication: \\ 
publication: August 2013 & August 2013 until July 2023. \\
until July 2023. & \\
\hline
- Is a journal or conference & - Review, book chapter, book, short \\
paper. & survey, editorial, retracted, conference \\
& review, data paper, note. \\
\hline
- Written in English. & - Other languages. \\ 
\hline
\end{tabular}
\caption{Inclusion and Exclusion Criteria for Paper Selection}
\label{tab:inclusionexclusion}
\end{table}

The collected papers from the initial search were screened according to the preset inclusion and exclusion criteria (Table \ref{tab:inclusionexclusion}). The paper selection process consisted of two phases. First, based on the inclusion and exclusion criteria, the papers were independently screened by two researchers through title and abstract screening (L.S \& I.G). The papers selected in this phase were then assessed through full-text screening (L.S. \& J.F.). The authors cross-checked the selection results and resolved any disagreement on the selection decisions. All disagreements in either the first or the second phase were resolved by consensus, and a third researcher (S.P.) was consulted to finalize the decision. The process is depicted in Fig. \ref{fig:flowchart}.

The inclusion and exclusion criteria were formulated by the authors to select relevant papers effectively. The documents underwent analysis to investigate diverse \ac{ml} approaches related to the energy efficiency and sustainability of \ac{iot} networks.

Both journal and conference papers written in English and within the scope of the research questions were included. Commercial papers, letters to the editor, ebooks, books, posters, and PhD dissertations were excluded. Papers were further excluded if 
\begin{itemize}
    \item they were review papers,
    \item they solely focused on the consideration of energy efficiency within \ac{ml} models,
    \item the energy efficiency was present in the hardware but not integrated into an \ac{iot} network,
    \item the technologies used were applicable to \ac{iot} but not integrated into an \ac{iot} network (e.g., UAVs, 5G, 6G, Mobile Edge Computing, etc.), or
    \item performance was not evaluated on hardware.
\end{itemize}
The authors deemed that the aim of the review was the integration of edge \ac{ml}, \ac{iot} networks and energy efficiency with the goal of enhancing sustainability.

Relevant data was extracted from the selected publications for further analysis:

\begin{itemize}
    \item Author list,
    \item Titles and abstracts,
    \item Year of publication, 
    \item Associated database, 
    \item \ac{iot} network application, \ac{iot} aspect optimized,
    \item \ac{ml} model(s) and metric used, sustainability gain achieved,
    \item \ac{iot} network specifications: communication technology, number and type of sensors used, hardware deployed
\end{itemize}

\subsection{Risk of Bias and Quality Assessment}

This systematic literature review adhered to the PRISMA guidelines for screening and selecting relevant literature. However, certain limitations must be acknowledged, primarily related to potential biases. The choice of keywords for the initial query search may introduce bias, and the subjectivity in defining eligibility criteria poses an additional risk. The reliance on a single database, Scopus, may limit the comprehensiveness of the literature search. Despite these considerations, the authors followed the best possible criteria in line with PRISMA guidelines. Independent selection processes and disagreement resolution techniques were employed to enhance transparency and objectivity. Additionally, a quality assessment system, established through consensus among authors, ensured the inclusion of high-quality publications with substantial contributions. This system was based on a checklist of the following criteria:

\begin{itemize}
    \item Are the methods used clearly defined and applied? 
    \item Are the methods applied successfully and correctly? 
    \item Are accuracy values and efficiency/confidence levels reported?
    \item Do the contributions outweigh the limitations of the study?
\end{itemize}

\subsection{Characteristics of selected papers}

The search process (Fig. \ref{fig:flowchart}) yielded a total of 5827 articles. After removing duplicates, 5820 papers remained. Of these, 5478 studies were excluded during the title and abstract screening for not meeting the inclusion criteria. Among the 342 studies that underwent full-text screening, 332 were found to not meet the full inclusion criteria and, therefore, further excluded. Ultimately, 10 studies were selected for the current review, as summarized in the following sections. The selection process took eleven months to complete.

The final selection for the systematic literature review comprised 9 journal papers \cite{xu_co-scheduling_2023, han_novel_2022, azar_energy_2019, kumari_energy_2022, gloria_autonomous_2021, sheth_eaps_2022, panda_energy-efficient_2023, savaglio_lightweight_2019, xiao_reinforcement_2023} and one conference paper \cite{lim_camthings_2018}. There was one paper each from China \cite{xiao_reinforcement_2023}, Italy \cite{savaglio_lightweight_2019}, the USA \cite{xu_co-scheduling_2023}, France \cite{azar_energy_2019}, Portugal \cite{gloria_autonomous_2021}, Canada \cite{panda_energy-efficient_2023}, Australia \cite{sheth_eaps_2022}, and India \cite{kumari_energy_2022}, and 2 papers from South Korea \cite{han_novel_2022, lim_camthings_2018}. These publications span the period from 2018 to 2023, with one publication in 2018 \cite{lim_camthings_2018} and 2021 \cite{gloria_autonomous_2021}, two in 2019 \cite{azar_energy_2019, savaglio_lightweight_2019} and three each in 2022 \cite{han_novel_2022, kumari_energy_2022, sheth_eaps_2022} and 2023 \cite{xu_co-scheduling_2023, panda_energy-efficient_2023, xiao_reinforcement_2023}.

\section{Findings}\label{results}

\begin{figure}
    \centering
    \includegraphics[width=0.5\linewidth]{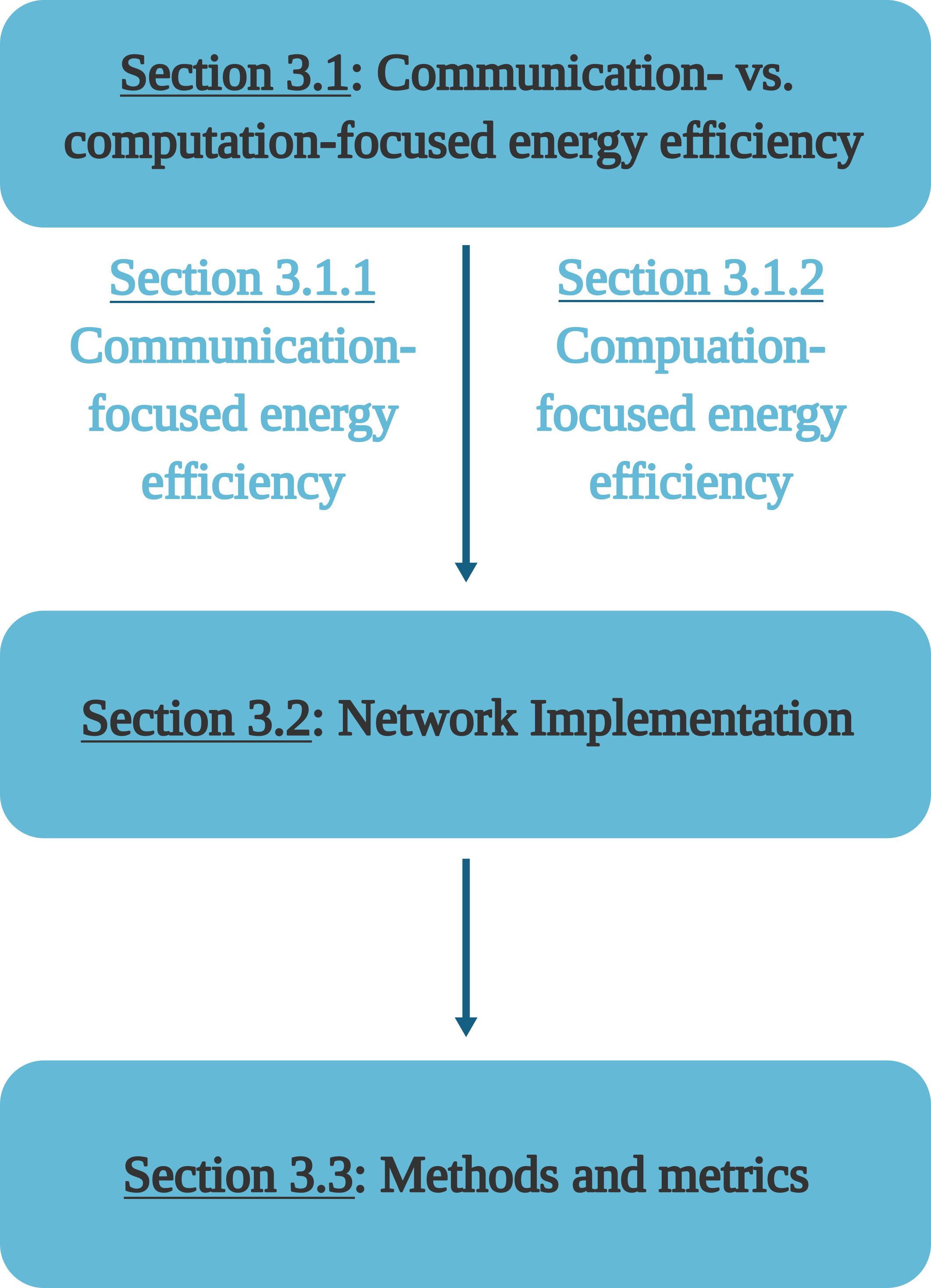}
    \caption{Layout of the findings section.}
    \label{fig:results_layout}
\end{figure}

To answer our research questions, we look into how the literature approached making \ac{iot} networks more sustainable.  The layout of this section is shown in Fig. \ref{fig:results_layout}. We first elucidate different aspects of an \ac{iot} network upon which sustainability can be improved. We differentiate between communication- and computation-focused approaches. Afterward, as we are interested in actual hardware implementations, we delve into the different \ac{iot} networks implemented in the selected papers, their specifications, and which kind of \ac{iot} application they execute. We finish by examining the different \ac{ml} methods used as well as metrics leveraged to measure the sustainability of an \ac{iot} network.

\subsection{Communication- vs. computation-focused energy efficiency}
To reduce the energy consumption of \ac{iot} networks, one can focus on the communication side of the network or on the computation level of each node in the network, respectively. Table \ref{tab:aspects} gives an overview of the network aspects in which sustainability was improved in the included papers, if its focus is on the communication or computation part, and how many papers looked into that specific aspect. It can be seen that the range of network design parameters that were researched is broad, tackling computation and communication subsystems.

\begin{table}[]
    \centering
    \begin{tabular}{|l|l|c|c|}
        \hline
        \textbf{Network aspect} & \textbf{Focus} & \textbf{Nr. of} & \textbf{Ref.} \\
        && \textbf{papers} & \\
        \hline
        Transmission period control & Communication & 1 & \cite{han_novel_2022} \\
        \hline
        Data compression & Communication & 2 & \cite{azar_energy_2019, kumari_energy_2022} \\
        \hline
        Sleep Scheduling & Communication & 1 & \cite{sheth_eaps_2022} \\
        \hline
        Communication protocol & Communication & 2 & \cite{gloria_autonomous_2021, savaglio_lightweight_2019} \\
        \hline
        Transmission selection & Communication & 1 & \cite{lim_camthings_2018} \\
        \hline
        Dynamic Voltage and & Computation & 1 & \cite{panda_energy-efficient_2023} \\
        Frequency Scaling &&& \\
        \hline
        Computation offloading & Computation & 2 & \cite{xiao_reinforcement_2023, lim_camthings_2018} \\
        \hline
        Co-scheduling of & Computation & 1 & \cite{xu_co-scheduling_2023} \\
        computational resources &&& \\
        \hline
    \end{tabular}
    \caption{Aspect of the network tackled to enhance sustainability, if this aspect focuses on the communication part of the network or on computation and how many papers dealt with this aspect.}
    \label{tab:aspects}
\end{table}

\subsubsection{Communication-focused energy efficiency}

We can further differentiate between network operation optimization and data transmission reduction when delving into communication-focused approaches. In the following, we start with covering papers that work on optimizing the network operation and then move on to those dealing with reducing the amount of data transmitted.

\paragraph{\underline {Network operation optimization}}

We find three aspects covered in the included papers, aiming to optimize the network operation: dynamically adjusting the data transmission period, tuning the duty cycle for sleep scheduling, and selecting the communication protocol. 

The authors in \cite{han_novel_2022} suggested increasing the sleep duration between each data transmission to transmit less often and, therefore, decreasing the energy consumption. This, however, entails that some data does not get transmitted at all. They dealt with this problem by introducing data imputation at the server. For this, they developed a \ac{dnn} consisting of a \ac{bilstm}, a \ac{cnn} and a simple one-layer \ac{ann}. This model takes the transmitted data and the data transmission period as input and, based on this, predicts the imputation accuracy. Using this result, they formulated an optimization problem to minimize the energy consumption of the \ac{iot} sensors and maximize the imputation accuracy. With their method, they achieved an energy consumption reduction of 18.23\% on the device.

The work in \cite{sheth_eaps_2022} is dedicated to improving sleep scheduling. In particular, a range of \ac{ml} methods were used to predict downlink packets so that the nodes only wake up when they expect data arrival. In this paper, different regression methods were compared, including \ac{rf}, \ac{gb}, \ac{et}, \ac{hb}, \ac{mlp} and \ac{lstm}. Finally, the \ac{lstm} model was implemented since it yielded the highest prediction accuracy. The authors of this work also compared their method to two power-saving modes, in two different scenarios: one using edge computing and one taking the computation of their application to the cloud. Their results are stated in Table \ref{tab:results_for_sheth}.
\begin{table}[h!]
    \centering
    \begin{tabular}{|l|l|c|}
        \hline
        \textbf{Power saving mode} & \textbf{Scenario} & \textbf{Energy efficiency} \\
        && \textbf{improvement} \\
        \hline
        PSM & Edge computing & 37\% \\
         & Cloud computing & 46\% \\
        \hline
        APSM & Edge computing & 6\% \\
         & Cloud computing & 26\% \\
        \hline
    \end{tabular}
    \caption{Energy efficiency improvement results from \cite{sheth_eaps_2022} for the normal power save mode (PSM) and the adaptive power save mode (APSM), executed in an edge computing scenario and in a cloud computing scenario.}
    \label{tab:results_for_sheth}
\end{table}

Another field of interest to improve the sustainability of a network's communication is the choice of the communication protocol. Two papers proposed a solution in this regard. The work in \cite{gloria_autonomous_2021} suggested self-configurable \ac{iot} nodes that choose the communication protocol and transmission power based on energy consumption and signal quality prediction. The protocols to choose from were ESP-Now \cite{esp-now}, \ac{ble} \cite{ble}, LoRa \cite{lorawan}, and ZigBee \cite{zigbee}. To determine the best prediction method, the work compared \ac{lr}, \ac{dt}, \ac{rf}, \ac{mlp} and \ac{svm}. As the \ac{rf} yielded the best prediction results, this was the model deployed in the network. Their method achieved up to 68\% energy consumption reduction with only 7\% of loss in network quality. While this approach used existing protocols and switched between them efficiently, \cite{savaglio_lightweight_2019} introduced a by-design energy-efficient MAC protocol. They used \ac{rl} to adjust the duty cycle of the node. In this way, they prolonged the node's lifetime by 4.5 times compared to the conventional CSMA-CA MAC with a duty cycle at 60\% and even by 26 times compared to the same conventional protocol with a duty cycle at 100\%.

\paragraph{\underline {Data transmission reduction}}

Two different approaches can be identified to reduce data transmissions. One approach is compressing the data before sending it to reduce the amount of transmitted bits and, therefore, save energy. A second approach involves selecting which data to transmit instead of blindly transmitting everything.

For the first approach, the authors of \cite{kumari_energy_2022} used an \ac{lstm}-based \ac{ae} to learn to compress and decompress the data to be sent. They leveraged optimization methods to find the optimal compression size, and in addition, they formulated another optimization problem to find the optimal data transmission speed. This problem was then solved with a conventional optimization method. They compared their method against existing approaches and demonstrated that they achieved lower energy consumption; however, no exact numbers were provided. The authors of \cite{azar_energy_2019} used traditional \ac{ml} methods, namely \ac{lcf} and \ac{qcf}, as part of the compression algorithm. Precisely, they only compressed unpredictable data points. Those data points predicted by the preceding \ac{ml} methods are discarded and predicted using the same method by the receiver. Using this method, they achieved an increase of 27\% of the device's lifetime after four hours.

The second approach is pursued in \cite{lim_camthings_2018}. The authors proposed to use a smart camera to train a \ac{dnn} to determine if a captured image is of interest or not, and to only transmit the image if it is classified as being of interest. This method decreased the energy consumption by 41\%. The specific \ac{dnn} architecture is, however, not specified in the paper. 

\subsubsection{Computation-focused energy efficiency}

Three aspects can be identified to enhance energy efficiency from the computational side. The work in \cite{panda_energy-efficient_2023} proposed the use of \ac{drl} to learn the optimal dynamic voltage and frequency scaling for local computation as well as the optimal data distribution for data offloading. It compared energy savings on three different \ac{iot} devices to two different Linux governors, which determines the \ac{cpu} utilization. The results are given in Table \ref{tab:results_for_panda}.
\begin{table}[h!]
    \centering
    \begin{tabular}{|l|l|c|}
        \hline
        \textbf{\ac{iot} device} & \textbf{Governor} & \textbf{Energy saving} \\
        \hline
        Linux laptop & Ondemand & 3.25\% - 5.78\% \\
         & Conservative & 9.9\% - 10.35\% \\
        \hline
        Nano 2 GB & Ondemand & 4.07\% - 7.72\% \\
         & Conservative & 7.13\% - 9.89\% \\
        \hline
    \end{tabular}
    \caption{Energy saving results from \cite{panda_energy-efficient_2023} for different \ac{iot} devices and CPU utilization governors. They also tested on a Raspberry Pi as an \ac{iot} device but stated that the energy savings there were negligible.}
    \label{tab:results_for_panda}
\end{table}

While \cite{lim_camthings_2018} offloaded data to reduce the computational load on the edge device, the research in \cite{xiao_reinforcement_2023} offloaded parts of \ac{dnn}s to divide computation between a local device and an edge server. It trained an \ac{rl} model and a \ac{drl} model to find the optimal \ac{dnn} model partition point. The authors stated that with the \ac{rl} model, the energy consumption of the local device was reduced by 13.9\% whereas, with the \ac{drl} model, this result was improved to a reduction of 41.8\%. However, the work mentioned that the edge server's energy consumption increased but did not elaborate further.

The offloading methods dealt with executing certain computations on a mobile device or on a server. Another way of optimizing computational load is to co-schedule computational resources, i.e., efficiently use computational resources on heterogeneous hardware. To achieve this, \cite{xu_co-scheduling_2023} leveraged \ac{drl} to co-schedule the use of \ac{cpu} and \ac{gpu}. They trained two models: one basic model to jointly optimize latency and throughput and one energy-aware model that optimizes these two factors and energy consumption. They compared their models to \ac{cpu}-only usage, \ac{gpu}-only usage, and a round-robin way of distributing computational load between \ac{cpu} and \ac{gpu}. Compared to \ac{cpu}-only, their basic model improved average energy consumption per task by 42.24\%, whereas their energy-aware model further increased this number by 4.09\%. They did not state numbers for the comparisons against \ac{gpu}-only and round-robin. However, from their results, it can be seen that both the energy-aware model and the basic model provided higher energy efficiency than the other baselines.

\subsection{Network implementation}

\begin{figure}
    \centering
    \includegraphics[width=0.9\linewidth]{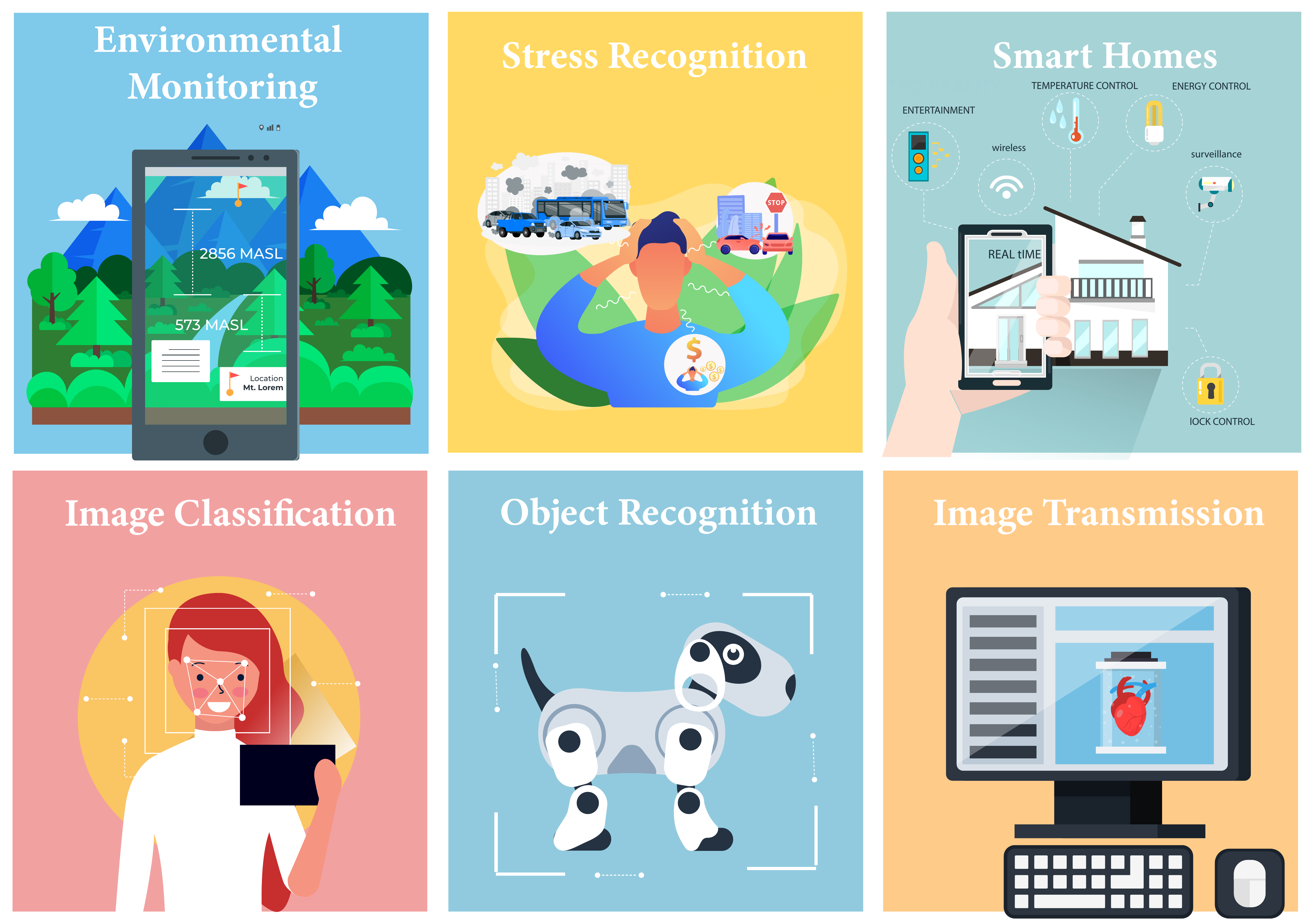}
    \caption{\ac{iot} applications in the included papers.}
    \label{fig:apps}
\end{figure}

\begin{figure}
    \centering
    \includegraphics[width=0.9\linewidth]{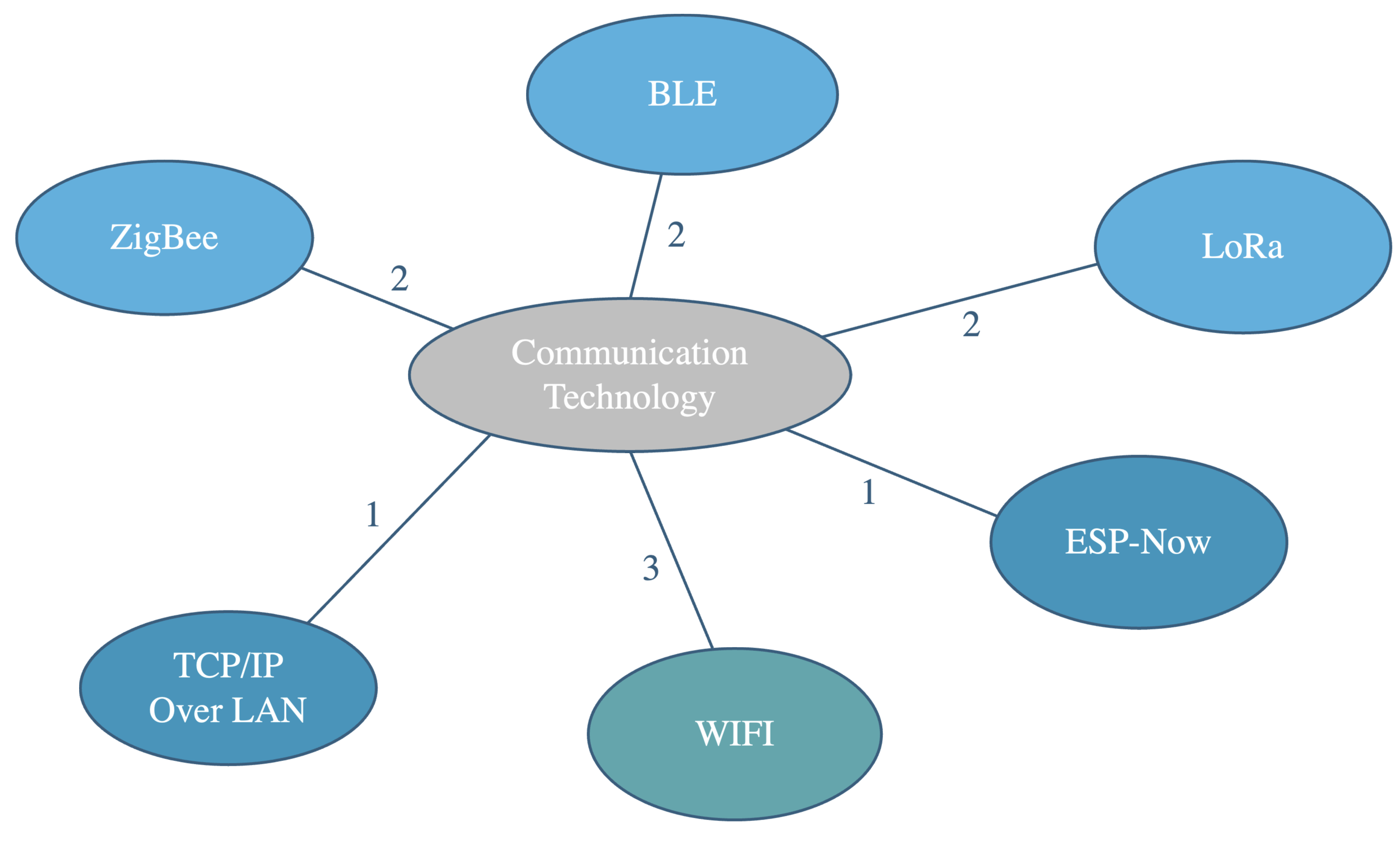}
    \caption{Different communication technologies and their usage count in the selected papers.}
    \label{fig:comm_techs}
\end{figure}

So far, we discussed how the literature has improved the sustainability of \ac{iot} networks. In this subsection, we will elaborate on the \ac{iot} network implementation in terms of scope, including network specifications and their \ac{iot} application. Fig. \ref{fig:apps} visualizes the different \ac{iot} applications. We find that three papers implemented full networks consisting of more than one point-to-point connection, while seven papers provided prototypes with no more than two connected devices to support their findings. We will now elaborate on the number of nodes, type of sensors, communication technology and hardware used, as well as the use the network served for. An overview of the communication technologies used is shown in Fig. \ref{fig:comm_techs}.

\subsubsection{Full network implementation}

\begin{table}[]
    \centering
   \begin{tabular}{|l|c|l|l|c|}
        \hline
        \textbf{Application} & \textbf{Nr.} & \textbf{Comm.} & \textbf{Hardware} & \textbf{Ref.} \\
        & \textbf{nodes} & \textbf{techn.} && \\
        \hline
        Dummy data & 35 & ESP-Now \cite{esp-now}, & ESP32-WROOM-32 \cite{espressif} & \cite{gloria_autonomous_2021} \\
        transmission && BLE \cite{ble}, && \\
        && LoRa \cite{lorawan}, && \\
        && ZigBee \cite{zigbee} && \\
        \hline
        Dummy data & 7 & ZigBee \cite{zigbee} & TelosB \cite{telosb} & \cite{savaglio_lightweight_2019} \\
        transmission &&&& \\
        \hline
        Dummy data & 4 & WiFi & Cypress CYW43907 \cite{cypress}, & \cite{sheth_eaps_2022} \\
        transmission &&& Raspberry Pi 4 \cite{rbp4}& \\
        \hline
    \end{tabular}
    \caption{Implemented full \ac{iot} networks in the included papers. No actual sensed data has been used, and therefore no sensor deployed.}
     \label{tab:network_specs_full}
\end{table}

Only a few works provided the implementation of a full network containing multiple nodes, summarized in Table \ref{tab:network_specs_full}. The biggest scope in the considered works was the network in \cite{gloria_autonomous_2021}, which consisted of 35 nodes and one gateway in an urban environment, spanning an area of 36ha. This network transmitted dummy data, generic to many applications for \ac{iot} networks. In this work, the authors switched between communication technologies, including four in total, namely ESP-Now \cite{esp-now}, \ac{ble} \cite{ble}, LoRa \cite{lorawan}, and ZigBee \cite{zigbee}. For their nodes, they used Espressif ESP31-WROOM-32 nodes \cite{espressif}. The second largest \ac{iot} network in the included work, implemented in \cite{savaglio_lightweight_2019}, also used ZigBee for communication. Seven TelosB nodes \cite{telosb} were implemented in their work, which similarly transmitted dummy data. The network in \cite{sheth_eaps_2022} featured four Cypress CYW43907 nodes \cite{cypress} and four Raspberry Pi edge devices. It was set in a residential area and transmitted, like the other networks, dummy data, but in their network, WiFi was being used for communication.

\subsubsection{Prototype implementation}

\begin{table}[]
    \centering
   \begin{tabular}{|l|c|l|l|l|c|}
        \hline
        \textbf{Application} & \textbf{Nr.} & \textbf{Type of} & \textbf{Comm.} & \textbf{Hardware} & \textbf{Ref.} \\
        & \textbf{nodes} & \textbf{sensors} & \textbf{techn.} && \\
        \hline
        Energy & 6 & Smart & LoRa \cite{lorawan} & DDS238-4W & \cite{kumari_energy_2022} \\
        consumption && meter && \cite{smartmeter}, RFM95W- & \\
        data transmis- && (energy && 868S2 \cite{loranode}, & \\ 
        sion in smart && consump- && Raspberry Pi 3 & \\
        home && tion) && \cite{rbp3}, Drigano & \\
        &&&& LG01-SIOT \cite{loragateway} & \\
        \hline
        Image & 2 & - & WiFi & Raspberry Pi 4 & \cite{xiao_reinforcement_2023} \\
        classification &&&& \cite{rbp4} & \\
        \hline
        Environmental & 1 & Tempera- & WiFi & Arduino Wemos & \cite{han_novel_2022} \\
        sensor data && ture, && D1 mini \cite{wemosmini}, & \\
        transmission && humidity, && Raspberry Pi 4 & \\
        && dust && \cite{rbp4} & \\
        \hline
        Stress rate & 1 & - & BLE \cite{ble} & Polar M600 & \cite{azar_energy_2019} \\
        classification &&&& wearable \cite{polar} & \\
        in car drivers &&&&& \\
        \hline
        Image & 1 & Image & BLE \cite{ble} & OV7725 \cite{imagesensor}, & \cite{lim_camthings_2018} \\
        transmission && sensor && Nordic nRF528 & \\
        &&&& 40 \cite{nrf}, Rasp- & \\
        &&&& berry Pi 3 \cite{rbp3} & \\
        \hline
        Image & 1 & - & TCP/IP & Raspberry Pi 4 & \cite{panda_energy-efficient_2023} \\
        classification &&& over LAN & \cite{rbp4}, Jetson & \\
        &&&& Nano 2 GB \cite{jetson}, & \\
        &&&& Linux laptop & \\
        &&&& with Intel CPU & \\
        &&&& \cite{intel}, Lenovo & \\
        &&&& Legion 5i laptop & \\
        &&&& \cite{lenovo} & \\
        \hline
        Facial expres- & 1 & - & - & Google Pixel 2 & \cite{xu_co-scheduling_2023} \\
        sion classifica- &&&& XL \cite{googlepixel} & \\
        tion and facial &&&&& \\
        landmarks &&&&& \\
        detection &&&&& \\
        \hline
    \end{tabular}
    \caption{Implemented prototypes in the included papers. If no sensor is stated, no actual sensed data has been used. Further, for the work of \cite{xu_co-scheduling_2023}, the communication technology is unimportant, so none is given.}
     \label{tab:network_specs_proto}
\end{table}

Most of the considered papers in the literature tested their methods on prototypes, with a maximum of two devices communicating with each other. We give an overview in Table \ref{tab:network_specs_proto}. Using WiFi, \cite{xiao_reinforcement_2023} used two Raspberry Pi 4's \cite{rbp4} as mobile devices and two edge servers to test their model for image classification. An existing dataset was used instead of images captured by the mobile devices. The authors in \cite{azar_energy_2019} tackled a very specific \ac{iot} task. They provided a prototype of a wearable to classify the stress rate of car drivers. More specifically, they used the Polar M600 wearable \cite{polar} and transmitted via \ac{ble} \cite{ble}. They first recorded signals from a driver, collected a data set like this, and then deployed it on the wearable to test their method. The results in \cite{panda_energy-efficient_2023} verified the efficacy of their method for the \ac{iot} application of image classification. In \cite{panda_energy-efficient_2023}, the authors used the well-known MNIST dataset \cite{lecun1998gradient} as dummy data for that and deployed it on a set of three different \ac{iot} devices. Those devices started small with a Raspberry Pi 4 \cite{rbp4}, getting more powerful over a Jetson Nano 2 GB \cite{jetson} to a Linux laptop with Intel \ac{cpu} \cite{intel}. All devices sent their data to a Lenovo Legion 5i laptop \cite{lenovo}, which served as an edge server. They only specified that they use TCP/IP over LAN for communication technology. The work in \cite{xu_co-scheduling_2023} did not entail communication. Their method was solely based on computation on the local device. Therefore, sending data for testing was not needed. As \ac{iot} device, they used the Google Pixel 2 XL \cite{googlepixel}, and the used application was facial expression classification and facial landmarks detection.

In \cite{xu_co-scheduling_2023, azar_energy_2019, gloria_autonomous_2021, sheth_eaps_2022, panda_energy-efficient_2023, savaglio_lightweight_2019, xiao_reinforcement_2023}, no sensors were deployed, and hence, no actual sensor readings were being used. While some of the works provided prototypes, they did use sensors. The authors of \cite{kumari_energy_2022} deployed DDS238-4W single-phase smart meters \cite{smartmeter} to measure the energy consumption of up to 12 appliances each in six households. The smart meter transmitted its readings using LoRa \cite{lorawan} to the Raspberry Pi 3 edge device \cite{rbp3}, which had an RFM95W-868S2 LoRa node \cite{loranode} attached to it. This node then further transmitted to a Drigano LG01-SIOT LoRa Gateway \cite{loragateway}. These readings were conducted every 20 seconds for 240 hours. In \cite{han_novel_2022}, the Arduino Wemos D1 mini sensor \cite{wemosmini} gathered information about temperature, humidity, and dust and transmitted it as an environmental monitoring application via WiFi to a Raspberry Pi 4B edge device \cite{rbp4}. The sensors were implemented in an office, and their experiment spanned 6 days. Sensor readings were performed at an interval of 10 seconds. It should be emphasized that both works tested their method on actual hardware in a real setting, coming very close to an actual application setting and, hence, trustworthy performance results. Last but not least, \cite{lim_camthings_2018} used an OV7725 image sensor \cite{imagesensor} to capture images. Connected to a Nordic nRF52840 node \cite{nrf}, these images were being transmitted through \ac{ble} \cite{ble} to a Raspberry Pi 3 gateway \cite{rbp3}. Information about the prototype's location, the experiment's time span, or amount of pictures taken were not given in the paper.


\subsection{Methods and metrics}

Finally, we examine the \ac{ml} methods and metrics that are being used in the included papers. For the \ac{ml} methods part, given the recent trend in using \ac{dl}, we specifically focus on the comparison between traditional \ac{ml} methods' and \ac{dl} methods' usage. From Table \ref{tab:ml_methods}, we can see that a variety of traditional \ac{ml} methods have been implemented, while only a few different \ac{dl} methods are considered.
\begin{table}[]
    \centering
    \begin{tabular}{|l|c|c|}
        \hline
        \textbf{\ac{ml} method} & \textbf{Nr. papers} & \textbf{Ref.} \\
        \hline
        \hline
        \ac{dt} & 1 & \cite{gloria_autonomous_2021} \\
        \hline
        \ac{et} & 1 & \cite{sheth_eaps_2022} \\
        \hline
        \ac{gb} & 1 & \cite{sheth_eaps_2022} \\
        \hline
        \ac{hb} & 1 & \cite{sheth_eaps_2022} \\
        \hline
        \ac{lcf} & 1 & \cite{azar_energy_2019} \\
        \hline
        \ac{lr} & 1 & \cite{gloria_autonomous_2021} \\
        \hline
        \ac{qcf} & 1 & \cite{sheth_eaps_2022} \\
        \hline
        \ac{rf} & 2 & \cite{sheth_eaps_2022, gloria_autonomous_2021} \\
        \hline
        \ac{rl} & 2 & \cite{savaglio_lightweight_2019, xiao_reinforcement_2023} \\
        \hline
        \ac{svm} & 1 & \cite{gloria_autonomous_2021} \\
        \hline
        \hline
        \ac{ann} & 3 & \cite{han_novel_2022, sheth_eaps_2022, gloria_autonomous_2021} \\
        \hline
        \ac{cnn} & 1 & \cite{han_novel_2022} \\
        \hline
        \ac{drl} & 3 & \cite{panda_energy-efficient_2023, xu_co-scheduling_2023, xiao_reinforcement_2023} \\
        \hline
        \ac{lstm} & 3 & \cite{han_novel_2022, kumari_energy_2022, sheth_eaps_2022} \\
        \hline
    \end{tabular}
    \caption{Overview of implemented \ac{ml} methods in the included papers and how many papers used the corresponding method. The first part covers the traditional \ac{ml} methods used, while the second part lists the \ac{dl} methods used. In "\ac{ann}", we summarize classical \ac{mlp}s and single-layer \ac{ann}s. Note that \cite{lim_camthings_2018} only stated that they use a \ac{dnn} but did not specify the architecture, so it is not listed here.}
    \label{tab:ml_methods}
\end{table}
However, each \ac{dl} method, except the \ac{cnn}, has been implemented in three papers, whereas the traditional methods are mainly used in one paper each. Only \ac{rf} and \ac{rl} have been used in two papers. Further, \cite{sheth_eaps_2022} implemented \ac{rf}, \ac{gb}, \ac{et}, \ac{hb}, an \ac{mlp} and an \ac{lstm} to compare their performance. The \ac{lstm} model is the only method implemented in real devices, which emerged as the best model in their comparison. Also, \cite{gloria_autonomous_2021} implemented different methods to compare, being \ac{lr}, \ac{dt}, \ac{rf}, an \ac{mlp}, and a \ac{svm}. In their work, however, the traditional \ac{ml} method \ac{rf} exhibited the best performance and beat the \ac{mlp}, a \ac{dl} model. In \cite{xiao_reinforcement_2023}, the authors compared traditional \ac{rl} against \ac{drl} and found that with \ac{drl}, more energy could be saved. Considering this, the papers where the best-achieving model was a traditional \ac{ml} model amount to three, while \ac{dl} models were implemented in the other seven papers. Note, however, that only three papers compared different \ac{ml} methods.

As for the evaluation metrics used, we can differentiate three metrics to indicate the sustainability of a network. Table \ref{tab:metrics} shows that seven out of ten papers evaluated their methods by how much the normal energy consumption is being reduced. In \cite{kumari_energy_2022}, the authors presented their method's performance by the percentage of energy efficiency improvement instead, being the only paper from the selected ones that uses this metric. In \cite{azar_energy_2019} and \cite{savaglio_lightweight_2019}, the authors used the lifetime of the \ac{iot} devices as an indicator and stated how much it increased using their proposed method.

\begin{table}[]
    \centering
    \begin{tabular}{|l|c|c|}
        \hline
        \textbf{Evaluation metric} & \textbf{Nr. papers} & \textbf{References} \\
        \hline
        Percentage energy & 7 & \cite{han_novel_2022, kumari_energy_2022, panda_energy-efficient_2023, gloria_autonomous_2021, lim_camthings_2018, xu_co-scheduling_2023, xiao_reinforcement_2023} \\
        consumption reduction && \\
        \hline
        Percentage energy & 1 & \cite{kumari_energy_2022} \\
        efficiency improvement && \\
        \hline
        Percentage device's & 2 & \cite{azar_energy_2019, savaglio_lightweight_2019} \\
        lifetime increase && \\
        \hline
    \end{tabular}
    \caption{Overview of metrics used to evaluate the sustainability enhancement of an \ac{iot} network.}
    \label{tab:metrics}
\end{table}

As most papers used the percentage in energy consumption reduction to evaluate the performance of their proposed models, we further detail their results. An individual comparison between the papers' implemented \ac{ml} methods and the gain in energy consumption reduction thereby achieved is shown in Table \ref{tab:energy_consumption_reduction}. The highest energy consumption reduction can be found in \cite{gloria_autonomous_2021} with going up to 68\%. In their work, the authors leveraged an \ac{rf} to adapt the communication protocol. This energy consumption reduction, however, comes at the price of a 7\% loss in network quality. The next highest gain can be found in \cite{xu_co-scheduling_2023}, with an energy consumption reduction of 46.33\%. Their approach encompasses a \ac{drl} model to efficiently co-schedule the usage of \ac{cpu} and \ac{gpu}. Followed closely regarding energy consumption reduction are the works in \cite{xiao_reinforcement_2023} and \cite{lim_camthings_2018}. Both advocate computational offloading to decrease energy consumption. In \cite{xiao_reinforcement_2023}, the authors train a \ac{drl} model to partition the used \ac{dnn} model and offload the second partition to the edge server, while in \cite{lim_camthings_2018}, data is offloaded only if it is classified by a \ac{dnn} to be of interest. It is further stated in \cite{xiao_reinforcement_2023} that this offloading increases the energy consumption of the edge server, a statement which also holds for \cite{lim_camthings_2018}. Both do not further elaborate on this drawback. Less energy consumption reduction is achieved in \cite{han_novel_2022, kumari_energy_2022, panda_energy-efficient_2023}. Their approaches tackle transmission period control, data compression, and dynamic voltage and frequency scaling, respectively. These results indicate that optimizing the usage and smartly combining existing resources, such as switching between communication protocols or optimizing the usage of \ac{cpu} and \ac{gpu}, yield higher energy consumption reduction potential. Less energy consumption reduction seems achievable by reducing the amount of transmitted data or going to very specific hardware instances, like voltage and frequency scaling. Note, however, that there is a mixture of applications, hardware and communication technologies in the papers discussed, so no definitive statement can be made.

\begin{table}[]
    \centering
    \begin{tabular}{|c|c|c|}
        \hline
        \textbf{Method} & \textbf{Gain} & \textbf{Reference} \\
        \hline
        \ac{drl} & 46.33\% & \cite{xu_co-scheduling_2023} \\
        \hline
        \ac{bilstm} + \ac{cnn} + \ac{ann} & 18.23\% & \cite{han_novel_2022} \\
        \hline
        \ac{lstm}-based \ac{ae} & - & \cite{kumari_energy_2022} \\
        \hline
        \ac{rf} & Up to 68\% (with 7\% & \cite{gloria_autonomous_2021} \\
        & loss in network quality) & \\
        \hline
        \ac{drl} & Up to 10.35\% & \cite{panda_energy-efficient_2023} \\
        \hline
        \ac{drl} & 41.8\% locally, but & \cite{xiao_reinforcement_2023} \\
        & increase in edge server & \\
        & energy consumption & \\
        \hline
        \ac{dnn} & 41\% & \cite{lim_camthings_2018} \\
        \hline
    \end{tabular}
    \caption{Comparison of implemented \ac{ml} techniques and their achieved energy consumption reduction. \cite{kumari_energy_2022} did not state exact numbers in their paper. Note that the gains of \cite{gloria_autonomous_2021} and \cite{xiao_reinforcement_2023} come at the loss of network quality and energy consumption increase at the server's side, respectively.}
    \label{tab:energy_consumption_reduction}
\end{table}



\section{Discussion}\label{discussion}

\begin{figure}
    \centering
    \includegraphics[width=0.99\linewidth]{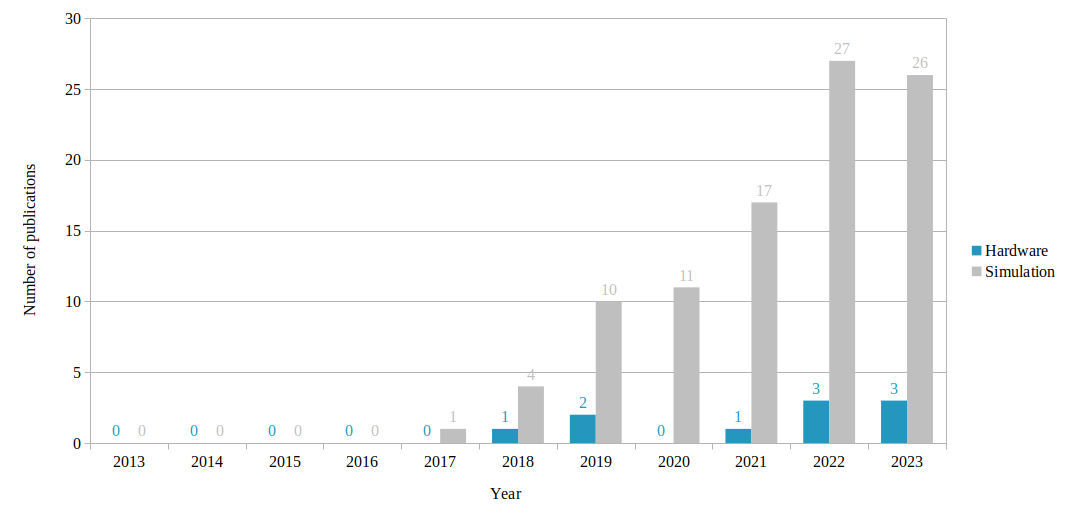}
    \caption{Distribution of selected papers over the included years: in cyan, final papers with hardware evaluation; in grey, papers that were evaluated with simulations but fulfilled all other inclusion criteria.}
    \label{fig:years}
\end{figure}

In this systematic literature review, we have explored state-of-the-art works that used \ac{ml} to enhance the sustainability of \ac{iot} networks. Of the 342 papers left after the abstract screening, 332 were excluded, and only 10 were included. Out of the 332 excluded papers, 96 papers fulfilled all criteria except actual hardware evaluation. This shows that almost tenfold of research evaluates their work on edge \ac{ml} for sustainable \ac{iot} with simulations rather than hardware. From the publication years of the included papers, we noticed that all fall in the second half of the selected period. Further, if we omit the very restrictive inclusion criterion of hardware-based evaluation, this trend still remains. The distribution for both is shown in Fig. \ref{fig:years}. This can be attributed to the fact that sustainability enhancement as a major goal in research, as well as a rapid increase in the development of \ac{dl} methods, has only evolved in recent years. For both to further arrive at niche areas like \ac{iot} networks could explain the lack of work before 2017.


Among the included papers, a wide range of \ac{iot} applications (Fig. \ref{fig:apps}) and aspects were investigated and optimized. In \cite{kim2023online}, the authors proposed to maximize the harvested energy of each \ac{iot} device while satisfying its data rate requirements. They did so utilizing an online \ac{drl} algorithm, which they ran distributed on each device, only using local channel state information. Going for the application of \ac{iiot}, \cite{wang2023dual} designed a \ac{drl} model using dual-attention for resource allocation. Another specific \ac{iot} field is the \ac{iomt}. The authors in \cite{wei2021intelligent} employed in this setting a \ac{drl} model as well and trained it to allocate channels so that Age of Information (AoI) and energy consumption are being optimized. Broadening again to \ac{iot} in general, a different approach for reducing energy consumption has been conducted in \cite{wang2020big}. They claimed that some of the collected data is untrustworthy due to harsh environments where sensor nodes are deployed and can be discarded before transmission. For that, they trained an \ac{svm} and further enhanced it with online learning, so that the model stays up-to-date upon employment. Focusing as well on the collected data, \cite{cao2020novel} suggests fusing data to reduce redundant information transmission. They, therefore, proposed an \ac{elm} and optimized it by a regular optimization algorithm. 

While simulations are a good starting point and give valuable insights, validating performance on actual hardware is crucial, as it might vary greatly from simulation performance \cite{caba2018testing}. From the included papers, which also performed hardware testing, we have seen various aspects of an \ac{iot} network being taken and investigated to improve energy efficiency. We split those aspects into communication-focused and computation-focused ones. While focusing on the communication part of the network is intuitive, with the trend of putting more computation on the edge of \ac{iot} networks \cite{gudnavar2023edge}, improving computational efficiency is of utmost importance as well. For the former one, we identified that transmission period control \cite{han_novel_2022}, data compression \cite{azar_energy_2019, kumari_energy_2022}, sleep scheduling \cite{sheth_eaps_2022}, the communication protocol itself \cite{gloria_autonomous_2021, savaglio_lightweight_2019}, and transmission selection \cite{lim_camthings_2018} have been taken to improve sustainability. For the latter one, we saw that dynamic voltage and frequency scaling \cite{panda_energy-efficient_2023}, computation offloading \cite{xiao_reinforcement_2023, lim_camthings_2018}, and co-scheduling of computational resources \cite{xu_co-scheduling_2023} have been deemed to have room for improvement via \ac{ml}. We have seen that more research has been conducted to enhance sustainability from the communication perspective \cite{han_novel_2022, azar_energy_2019, kumari_energy_2022, sheth_eaps_2022, gloria_autonomous_2021, savaglio_lightweight_2019, lim_camthings_2018} rather than the computational perspective \cite{panda_energy-efficient_2023, xiao_reinforcement_2023, lim_camthings_2018, xu_co-scheduling_2023}. Though this variety of network aspects showcases creativity in how to make \ac{iot} networks more sustainable, it also demonstrates that tackling a different area for improving sustainability is favored over improving existing areas. This is emphasized by the fact that baseline comparison happens mainly to the normal network operation \cite{xu_co-scheduling_2023, azar_energy_2019, lim_camthings_2018, sheth_eaps_2022, panda_energy-efficient_2023, savaglio_lightweight_2019, gloria_autonomous_2021}. Only \cite{han_novel_2022, kumari_energy_2022, xiao_reinforcement_2023} compare their methodology to existing works. Building on and outperforming existing work, like done so often in the \ac{ml} community as challenges with leaderboards and leading to rapid performance improvement \cite{donoho2024data}, seems to not be implemented yet.

Different approaches have been taken to improve energy efficiency, and we have seen a diversity of \ac{ml} methods, partially intertwined, to realize them. Those techniques range from classical and well-known \ac{ml} methods like \ac{dt}s and \ac{svm}s \cite{gloria_autonomous_2021} to less popular methods like \ac{hb} \cite{sheth_eaps_2022} and to booming \ac{dl} methods like \ac{drl} \cite{xu_co-scheduling_2023, panda_energy-efficient_2023, xiao_reinforcement_2023} and \ac{lstm}s \cite{han_novel_2022, kumari_energy_2022, sheth_eaps_2022}. A trend towards \ac{dl} models could be identified, whereas \ac{drl} \cite{xu_co-scheduling_2023, panda_energy-efficient_2023, xiao_reinforcement_2023} and \ac{lstm}s \cite{han_novel_2022, kumari_energy_2022, sheth_eaps_2022} took the lead in mainly used models. It should be highlighted that traditional \ac{ml} methods were investigated \cite{azar_energy_2019, gloria_autonomous_2021, sheth_eaps_2022, savaglio_lightweight_2019, xiao_reinforcement_2023} and in one instance, they outperformed the nowadays hyped \ac{dl} models \cite{gloria_autonomous_2021}. However, this statement has to be viewed with caution, as not necessarily the best fit \ac{dl} model or \ac{ml} model has been implemented for comparison.

We further investigated the implemented \ac{iot} networks and discovered that various hardware and \ac{iot} applications have been used to verify the proposed method's effectiveness. The \ac{iot} applications utilized dealt with either simply data transmission \cite{han_novel_2022, azar_energy_2019, kumari_energy_2022, gloria_autonomous_2021, savaglio_lightweight_2019, sheth_eaps_2022, lim_camthings_2018}, or was an image-based application, like object recognition \cite{xiao_reinforcement_2023, xu_co-scheduling_2023} or image classification \cite{panda_energy-efficient_2023}. To implement these applications, three papers leveraged a full \ac{iot} network in a for their purpose adequate environment, ranging from a total of 4 nodes to a total number of 35 nodes \cite{gloria_autonomous_2021, sheth_eaps_2022, savaglio_lightweight_2019}. They, however, only used dummy data to transmit instead of real sensor readings. The other seven papers provided prototypes to test their proposed method \cite{xu_co-scheduling_2023, han_novel_2022, azar_energy_2019, kumari_energy_2022, panda_energy-efficient_2023, xiao_reinforcement_2023, lim_camthings_2018}. Three papers started with simulations and then backed up their findings with prototype implementations \cite{kumari_energy_2022, savaglio_lightweight_2019, xiao_reinforcement_2023}. Communication technologies used were ESP-Now \cite{gloria_autonomous_2021}, \ac{ble} \cite{gloria_autonomous_2021, azar_energy_2019, lim_camthings_2018}, LoRa \cite{gloria_autonomous_2021, kumari_energy_2022}, ZigBee \cite{gloria_autonomous_2021, savaglio_lightweight_2019}, and WiFi \cite{sheth_eaps_2022, xiao_reinforcement_2023, han_novel_2022} (Fig. \ref{fig:comm_techs}). The data transmitted by the networks were either dummy sensor readings \cite{gloria_autonomous_2021, savaglio_lightweight_2019, sheth_eaps_2022}, images/sensed data loaded on the node beforehand \cite{xiao_reinforcement_2023, azar_energy_2019, panda_energy-efficient_2023, xu_co-scheduling_2023}, energy consumption readings from a smart meter \cite{kumari_energy_2022}, environmental sensor readings \cite{han_novel_2022}, or images captured by an image sensor \cite{lim_camthings_2018}. Distinguishing themselves from the rest are the last three works \cite{han_novel_2022, kumari_energy_2022, lim_camthings_2018}. They very closely replicate actual applications by using sensor readings on the device instead of dummy data and being set in realistic environments. In total, we can see that a multitude of different \ac{iot} applications and hardware were considered. We emphasize at this point again the value of executing \ac{ml} models on actual hardware, further strengthening trustworthiness by the usage of sensor readings on device, and evaluating their performance in this way rather than only through simulations. However, broadening the evaluation to several different \ac{iot} networks and showing that the proposed method gives the expected performance in different scenarios and environments would further increase credibility for deployment by other researchers or industry.

As for the performance metrics, all of them are concerned with energy, be it just reducing its consumption \cite{xu_co-scheduling_2023, han_novel_2022, gloria_autonomous_2021, sheth_eaps_2022, panda_energy-efficient_2023, xiao_reinforcement_2023, lim_camthings_2018}, improving efficiency \cite{kumari_energy_2022}, or measuring its impact by an increase in a device's lifetime \cite{azar_energy_2019, savaglio_lightweight_2019}. Despite them being all related, having different metrics makes it difficult to compare methods from different papers. But even if the same metric was being used throughout the literature, differences like \ac{iot} application, communication technology, hardware, or network size, among others, make it, in any case, practically impossible to compare findings from different papers \cite{gloria2017comparison}.

\subsection{Identified gaps}
From our findings, we identify the following gaps in the current literature:
\begin{itemize}
    \item Lack of evaluation in real \ac{iot} networks. Though 342 papers were selected after the full-text screening to match all inclusion criteria, excluding the implementation of actual hardware, the majority carried out simulations to evaluate the effectiveness of their model. Although simulations provide important insights and can be a first step when proposing a method to enhance an \ac{iot} network's sustainability, testing the method on actual hardware is necessary. There are two main reasons for this. First, simulation results often greatly differ from performance on real hardware \cite{caba2018testing}. Second, implementing and executing \ac{ml} models on resource-constrained hardware is a hurdle that might need to be overcome with modifications of the model, which then may lead to poorer performance than before. Further, only three offered a complete \ac{iot} network implementation from the ten included papers. The other seven papers provided prototypes consisting of two connected devices. Testing in a bigger \ac{iot} network consisting of multiple nodes and implemented in an environment where it would be implemented in practice is expected to uncover unseen issues and performance variations. However, the full network implementations only transmit dummy data, neglecting the effect the usage of actual sensor readings could have, i.e., their temporal and spatial dynamics.
    \item Lack of implementation of computation models in real hardware, along with a quantitative assessment of their energy and latency requirements. This involves the identification of the tradeoffs between the energy savings and other performance improvements (e.g., related to decision-making for network or protocol control) that are enabled by the use of \ac{ml} and the additional energy and latency burden due to its execution. This includes the investigation of \ac{ml} models in isolation and, subsequently, their integration into the whole \ac{iot} ecosystem, jointly considering communication and scheduling aspects.
    \item Lack of available datasets and code. In the spirit of open science, making datasets and code accessible would greatly help advance research further. Other researchers can build on existing code or work with existing datasets, making it additionally meaningful to compare methods against each other on the same dataset. The implemented work would be reproducible, aiding in actual implementation in the industry. Despite this, {\it none of the included papers} provided the data or code open-source, except those that use existing datasets that have been open-source before already, like MNIST. Falling in this category would also be to further make prototypes / implemented networks accessible. Collaboration between researchers could be enhanced and existing \ac{iot} infrastructure leveraged.
    \item Lack of anomaly management. All methods assume normal network behavior. This is not necessarily the case in real-life implementations. Anomalies in the environment can impact communication, which could lead to drastic quality drops for those methods, increasing energy efficiency through compromising network quality. Also, methods focusing on computational efficiency might lead to unreasonable performance when hardware anomalies in the computational unit of the sensors occur.
    \item Lack of comparison between different models and algorithms to enhance sustainability. Most papers propose an \ac{ml} method or \ac{dl} model architecture and then compare it to the normal network operation or standard energy-saving methods. To support the superiority of the proposed method, comparing it to other (\ac{ml}) methods is recommended. Ablation studies to ensure optimal hyperparameter settings, to which \ac{ml} methods are very sensitive, would also be appreciated to convince that the methods described are chosen in the best way possible for the given task.
\end{itemize}

\subsection{Directions for future research}

Taking the findings from our systematic literature review and the identified gaps into account, we suggest the following direction for future research:
\begin{itemize}
    \item Implementing more advanced \ac{dl} models. While \ac{drl} and \ac{lstm}s are powerful models from the \ac{dl} family, in recent years, other mechanisms have risen in the booming fields of image processing and natural language processing. To name the prominent attention mechanism, these kinds of recent advances in \ac{dl} have not been applied to the field of sustainable \ac{iot} yet. Though originating from image processing or natural language processing, those new methods are being successfully applied in many other research areas, making them promising to yield good performance for sustainable \ac{iot}. Along these lines, modern energy-efficient \ac{ml} designs, such as \ac{elm}s \cite{huang2006extreme}, reservoir neural networks \cite{lukovsevivcius2009reservoir} or even spiking neural networks and neuromorphic hardware \cite{tavanaei2019deep, young2019review} should be considered in addition to classical convolutional, recurrent networks or transformer architectures.
    \item The energy and latency cost of \ac{ml} and \ac{dl} models is still insufficiently explored. Analytical, semi-analytical or empirical models to estimate the energy consumption of neural network architectures would be highly valuable to implement optimization-based methods for two main reasons: to come up with highly energy efficient \ac{ml} architectures for \ac{iot} via network architecture search, and to optimize \ac{ml} models at runtime, e.g., via split computing or early exiting \cite{matsubara2022split}, in an effort to continuously minimizing their energy footprint. Therefore, the design of a metric to measure the energy cost of \ac{ml} and \ac{dl} models independent of the used hardware, which will be accepted as the metric to use for energy cost in the community, is of utmost importance to compare and assess research as well.
    \item Comprehensive frameworks, with solid validation on hardware, including neural network models and communication protocols, are still missing where the two are jointly optimized. This is the natural evolution of current network technology, which works following rigid protocol rules, into future ``intelligent'' networks with self-optimization and self-assessment capabilities. This objective is being discussed extensively \cite{6gwhitepaper}, and while simulation-based research can be found \cite{biason2017eccentric}, methods validated on hardware are still lacking in the \ac{iot} domain. 
    \item Open access to testbeds. Many more papers evaluate their methods on simulations rather than hardware due to a lack of financial means or expertise for building intricate testbeds for evaluation. Moreover, building new testbeds in each research group contradicts the sustainability principle, as many chips and hardware would be needed. Another crucial point is the time and effort it takes to build a testbed, which could be invested in further research instead. Therefore, we advocate using existing testbeds by making their usage also possible outside the research group implementing the testbed. This can be achieved either by collaboration or by taking it a step further, by making testbeds accessible and their collected datasets open access to the whole research community \cite{6gbricks}.
    \item Consideration of the existence of anomalies. To support the methods' robustness, showcasing that existing methods also work in the presence of anomalies is a field of interest. In case methods cease to work properly in the presence of anomalies, anomaly detection, and removal should be investigated. Anomalies can not only hinder the effectiveness of proposed methods to enhance sustainability, but they also increase energy consumption. Detecting and removing them is a gap yet to be filled.
    \item Verifying existing methods evaluated by simulations on actual hardware. As we already stated, the amount of research based on solely simulations outnumbers actual hardware implementations. Instead of pushing forward for new methods, verifying existing ones and making them therefore applicable in real \ac{iot} networks would carry a high value.
\end{itemize}

\subsection{Limitations}

The findings of this review should be considered in light of some limitations. Although the data source covered many scientific databases, they did not encompass all available literature, limiting the generalizability of the findings. The review specifically focused on actual hardware implementations, excluding many approaches to optimizing an \ac{iot} network's sustainability that were evaluated solely through simulations. Though the limitations of simulation-based results were discussed and their exclusion motivated, it must be noted that many simulations follow real-world-like assumptions, and their findings are meaningful and insightful for application on hardware. Collecting papers and reviewing them based on the quality of the simulations would be another critical contribution. Additionally, relevant works may have emerged during and beyond the specified data collection period, which spanned from August 2013 to July 2023. Finally, despite meticulous data extraction and analysis, the potential for bias remains.

\section{Conclusion}\label{conclusion}

Both \ac{iot} networks and \ac{ml} have experienced an explosion in usage in recent years. \ac{iot} networks are getting deployed in many different environments for monitoring and automation, while \ac{ml} aids in optimizing routines and finding solutions to problems traditional methods are not capable of solving. \ac{iot} networks, however, come at the expense of the environment. They consist of nodes using chips that must be manufactured and later disposed of. Further, their operation is energy-consuming. In this systematic literature review, we have looked into the application of \ac{ml} on the edge of \ac{iot} networks to increase their sustainability. We could differentiate between communication-focused and computation-focused approaches, where \ac{ml} has been used to optimize the network's communication or the computation on the \ac{iot} nodes. We have found that a variety of different aspects of an \ac{iot} network have been tackled to optimize by using \ac{ml}, as well as numerous different kinds of \ac{iot} networks have been implemented to validate the proposed \ac{ml} methodology. Based on this, we have identified gaps and proposed directions for future research.

\section*{Declaration of generative AI and AI-assisted technologies in the writing process}
During the preparation of this work, the authors used ChatGPT in order to reformulate the abstract and introduction for a better reading flow. After using this tool, the authors reviewed and edited the content as needed and take full responsibility for the content of the published article.





\begin{thebibliography}{10}
\expandafter\ifx\csname url\endcsname\relax
  \def\url#1{\texttt{#1}}\fi
\expandafter\ifx\csname urlprefix\endcsname\relax\def\urlprefix{URL }\fi
\expandafter\ifx\csname href\endcsname\relax
  \def\href#1#2{#2} \def\path#1{#1}\fi

\bibitem{prabhu_iot_2024}
H.~Prabhu, J.~Joseph, K.~K. Koushik, J.~Karthikeyan, M.~Deeksha, {IoT} and its potential for transforming industries, International Journal of Networks and Systems (2024).

\bibitem{prakash_green_2023}
R.~Prakash, D.~Singh, Green {Internet} of {Things} ({G-IOT}) for sustainable environment, EPRA International Journal of Multidisciplinary Research (2023).

\bibitem{raghavendar_robust_2023}
K.~Raghavendar, I.~Batra, A.~Malik, A robust resource allocation model for optimizing data skew and consumption rate in cloud-based {IoT} environments, Decision Analytics Journal (2023).

\bibitem{mukhamediev2022review}
R.~I. Mukhamediev, Y.~Popova, Y.~Kuchin, E.~Zaitseva, A.~Kalimoldayev, A.~Symagulov, V.~Levashenko, F.~Abdoldina, V.~Gopejenko, K.~Yakunin, et~al., Review of {Artificial} {Intelligence} and {Machine} {Learning} technologies: Classification, restrictions, opportunities and challenges, Mathematics (2022).

\bibitem{venigandla2022integrating}
K.~Venigandla, Integrating {RPA} with {AI} and {ML} for enhanced diagnostic accuracy in healthcare, Power System Technology (2022).

\bibitem{tatineni2024enhancing}
S.~Tatineni, A.~Mustyala, Enhancing financial security: {Data} {Science}'s role in risk management and fraud detection, ESP International Journal of Advancements in Computational Technology (2024).

\bibitem{chung2021applications}
S.-H. Chung, Applications of smart technologies in logistics and transport: A review, Transportation Research Part E: Logistics and Transportation Review (2021).

\bibitem{khayyam2020artificial}
H.~Khayyam, B.~Javadi, M.~Jalili, R.~N. Jazar, Artificial {Intelligence} and {Internet} of {Things} for autonomous vehicles, Nonlinear Approaches in Engineering Applications: Automotive Applications of Engineering Problems (2020).

\bibitem{averbuch2022applications}
T.~Averbuch, K.~Sullivan, A.~Sauer, M.~A. Mamas, A.~A. Voors, C.~P. Gale, M.~Metra, N.~Ravindra, H.~G. Van~Spall, Applications of {Artificial} {Intelligence} and {Machine} {Learning} in heart failure, European Heart Journal-Digital Health (2022).

\bibitem{alkhayyal2024recent}
M.~Alkhayyal, A.~Mostafa, Recent developments in {AI} and {ML} for {IoT}: A systematic literature review on {LoRaWAN} energy efficiency and performance optimization, Sensors (2024).

\bibitem{zhou2017machine}
L.~Zhou, S.~Pan, J.~Wang, A.~V. Vasilakos, Machine {Learning} on big data: Opportunities and challenges, Neurocomputing (2017).

\bibitem{yu2017survey}
W.~Yu, F.~Liang, X.~He, W.~G. Hatcher, C.~Lu, J.~Lin, X.~Yang, A survey on the {Edge} {Computing} for the {Internet} of {Things}, IEEE Access (2017).

\bibitem{ning2018green}
Z.~Ning, X.~Kong, F.~Xia, W.~Hou, X.~Wang, Green and sustainable {Cloud} of {Things}: Enabling collaborative {Edge} {Computing}, IEEE Communications Magazine (2018).

\bibitem{almutairi2024advancements}
R.~Almutairi, G.~Bergami, G.~Morgan, Advancements and challenges in {IoT} simulators: A comprehensive review, Sensors (2024).

\bibitem{tekin2023energy}
N.~Tekin, A.~Acar, A.~Aris, A.~S. Uluagac, V.~C. Gungor, Energy consumption of on-device {Machine} {Learning} models for {IoT} intrusion detection, Internet of Things (2023).

\bibitem{schuhmacher2023ecoble}
L.~Schuhmacher, S.~Pollin, H.~Sallouha, {ecoBLE}: A low-computation energy consumption prediction framework for {Bluetooth} {Low} {Energy}, International Conference on Embedded Wireless Systems and Networks (2023).

\bibitem{djigal2022machine}
H.~Djigal, J.~Xu, L.~Liu, Y.~Zhang, Machine and {Deep} {Learning} for resource allocation in multi-access {Edge} {Computing}: A survey, IEEE Communications Surveys \& Tutorials (2022).

\bibitem{sami2023forecasting}
M.~A. Sami, T.~A. Khan, Forecasting failure rate of {IoT} devices: A {Deep} {Learning} way to predictive maintenance, Computers and Electrical Engineering (2023).

\bibitem{merenda}
M.~Merenda, C.~Porcaro, D.~Iero, Edge {Machine} {Learning} for {AI}-enabled {IoT} devices: A review, Sensors (2020).

\bibitem{murshed}
M.~G.~S. Murshed, C.~Murphy, D.~Hou, N.~Khan, G.~Ananthanarayanan, F.~Hussain, Machine {Learning} at the network edge: A survey, ACM Computing Surveys (2021).

\bibitem{survey1}
N.~Charef, A.~{Ben Mnaouer}, M.~Aloqaily, O.~Bouachir, M.~Guizani, Artificial {Intelligence} implication on energy sustainability in {Internet} of {Things}: A survey, Information Processing \& Management (2023).

\bibitem{slr}
D.~Wang, D.~Zhong, A.~Souri, Energy management solutions in the {Internet} of {Things} applications: Technical analysis and new research directions, Cognitive Systems Research (2021).

\bibitem{comp_study}
A.~B. Guiloufi, S.~El~khediri, N.~Nasri, A.~Kachouri, A comparative study of energy efficient algorithms for {IoT} applications based on {WSN}s, Multimedia Tools and Applications (2023).

\bibitem{page2021prisma}
M.~J. Page, J.~E. McKenzie, P.~M. Bossuyt, I.~Boutron, T.~C. Hoffmann, C.~D. Mulrow, L.~Shamseer, J.~M. Tetzlaff, E.~A. Akl, S.~E. Brennan, et~al., The {PRISMA} 2020 statement: An updated guideline for reporting systematic reviews, British Medical Journal (2021).

\bibitem{SLRig}
I.~Gryech, C.~Asaad, M.~Ghogho, A.~Kobbane, Applications of {Machine} {Learning} \& {Internet} of {Things} for outdoor air pollution monitoring and prediction: A systematic literature review, Engineering Applications of Artificial Intelligence (2024).

\bibitem{SLRuv}
I.~Gryech, E.~Vinogradov, A.~Saboor, P.~S. Bithas, P.~T. Mathiopoulos, S.~Pollin, A systematic literature review on the role of {UAV}-enabled communications in advancing the {UN}’s sustainable development goals, Frontiers in Communications and Networks (2024).

\bibitem{li2022energy}
D.~Li, M.~Lan, Y.~Hu, Energy-saving service management technology of {Internet} of {Things} using {Edge} {Computing} and {Deep} {Learning}, Complex \& Intelligent Systems (2022).

\bibitem{chen2021energy}
X.~Chen, J.~Zhang, B.~Lin, Z.~Chen, K.~Wolter, G.~Min, Energy-efficient offloading for {DNN}-based smart {IoT} systems in cloud-edge environments, IEEE Transactions on Parallel and Distributed Systems (2021).

\bibitem{bebortta2021robust}
S.~Bebortta, A.~K. Singh, B.~Pati, D.~Senapati, A robust energy optimization and data reduction scheme for {IoT} based indoor environments using local processing framework, Journal of Network and Systems Management (2021).

\bibitem{cai2019iot}
Y.~Cai, A.~Genovese, V.~Piuri, F.~Scotti, M.~Siegel, {IoT}-based architectures for sensing and local data processing in ambient intelligence: Research and industrial trends, IEEE International Instrumentation and Measurement Technology Conference (2019).

\bibitem{gatti2019grey}
L.~Gatti, P.~Seele, L.~Rademacher, Grey zone in--greenwash out. a review of greenwashing research and implications for the voluntary-mandatory transition of {CSR}, International Journal of Corporate Social Responsibility (2019).

\bibitem{scopus}
Scopus. 2024 elsevier b.v., \\ https://www.scopus.com. [online] accessed on: September 2024.

\bibitem{Schotten}
M.~Schotten, M.~{El Aisati}, W.~Meester, S.~Steiginga, C.~Ross, A brief history of scopus: The world{\textquoteright}s largest abstract and citation database of scientific literature, Research Analytics (2017).

\bibitem{Rayyan}
M.~Ouzzani, H.~Hammady, Z.~Fedorowicz, A.~Elmagarmid, Rayyan — a web and mobile app for systematic reviews, Systematic Reviews (2016).

\bibitem{xu_co-scheduling_2023}
Z.~Xu, D.~Yang, C.~Yin, J.~Tang, Y.~Wang, G.~Xue, A co-scheduling framework for {DNN} models on mobile and edge devices with heterogeneous hardware, IEEE Transactions on Mobile Computing (2023).

\bibitem{han_novel_2022}
J.~Han, G.~H. Lee, J.~Lee, T.~Y. Kim, J.~K. Choi, A novel {Deep}-{Learning}-based robust data transmission period control framework in {IoT} {Edge} {Computing} system, IEEE Internet of Things Journal (2022).

\bibitem{azar_energy_2019}
J.~Azar, A.~Makhoul, M.~Barhamgi, R.~Couturier, An energy efficient {IoT} data compression approach for edge {Machine} {Learning}, Future Generation Computer Systems (2019).

\bibitem{kumari_energy_2022}
P.~Kumari, R.~Mishra, H.~P. Gupta, T.~Dutta, S.~K. Das, An energy efficient smart metering system using {Edge} {Computing} in {LoRa} network, IEEE Transactions on Sustainable Computing (2022).

\bibitem{gloria_autonomous_2021}
A.~F.~X. Glória, P.~J.~A. Sebastião, Autonomous configuration of communication systems for {IoT} smart nodes supported by {Machine} {Learning}, IEEE Access (2021).

\bibitem{sheth_eaps_2022}
J.~Sheth, C.~Miremadi, A.~Dezfouli, B.~Dezfouli, {EAPS}: Edge-assisted predictive sleep scheduling for 802.11 {IoT} stations, IEEE Systems Journal (2022).

\bibitem{panda_energy-efficient_2023}
S.~K. Panda, M.~Lin, T.~Zhou, Energy-efficient computation offloading with {DVFS} using {Deep} {Reinforcement} {Learning} for time-critical {IoT} applications in {Edge} {Computing}, IEEE Internet of Things Journal (2023).

\bibitem{savaglio_lightweight_2019}
C.~Savaglio, P.~Pace, G.~Aloi, A.~Liotta, G.~Fortino, Lightweight {Reinforcement} {Learning} for energy efficient communications in {Wireless} {Sensor} {Networks}, IEEE Access (2019).

\bibitem{xiao_reinforcement_2023}
Y.~Xiao, L.~Xiao, K.~Wan, H.~Yang, Y.~Zhang, Y.~Wu, Y.~Zhang, Reinforcement {Learning} based energy-efficient collaborative inference for mobile {Edge} {Computing}, IEEE Transactions on Communications (2023).

\bibitem{lim_camthings_2018}
J.~Lim, J.~Seo, Y.~Baek, {CamThings}: {IoT} camera with energy-efficient communication by {Edge} {Computing} based on {Deep} {Learning}, {International} {Telecommunication} {Networks} and {Applications} {Conference} (2018).

\bibitem{esp-now}
2021 {Espressif} {Systems} ({Shanghai}) {Co}., {Ltd}. {ESP-Now} overview. \\ https://www.espressif.com/en/solutions/low-power-solutions/esp-now. [online] accessed on: September 2024.

\bibitem{ble}
2017 {Bluetooth} {SIG}, {Inc}. {Bluetooth} {Low} {Energy}. \\ https://web.archive.org/web/20170310111443/https://www.bluetooth. com/what-is-bluetooth-technology/how-it-works/low-energy. [online] accessed on: April 2025.

\bibitem{lorawan}
2024 {LoRa} {Alliance}. what is {LoRaWAN}? \\ https://lora-alliance.org/about-lorawan. [online] accessed on: September 2024.

\bibitem{zigbee}
2015 {ZigBee} {Alliance}. {ZigBee} specifications. \\ https://zigbeealliance.org/wp-content/uploads/2019/11/docs-05-3474-21-0csg-zigbee-specification.pdf. [online] accessed on: September 2024.

\bibitem{espressif}
2023 {Espressif} {Systems} ({Shanghai}) {Co.}, {Ltd}. {ESP32-WROOM-32} datasheet. \\ https://www.espressif.com/sites/default/files/documentation/esp32-wroom-32\_datasheet\_en.pdf. [online] accessed on: July 2024.

\bibitem{telosb}
{CrossbowTechnology}, {Inc}. {TelosB} mote platform. \\ https://www.willow.co.uk/telosb\_datasheet.pdf. [online] accessed on: July 2024.

\bibitem{cypress}
Cypress {Semiconductor}. {CYW43907}: {IEEE} 802.11 a/b/g/n {SoC} with an embedded applications processor, \\ http://www.cypress.com/file/298236/download. [online] accessed on: September 2024.

\bibitem{rbp4}
Raspberry {Pi} {Ltd}. {Raspberry} {Pi} 4 model {B}, \\ https://datasheets.raspberrypi.com/rpi4/raspberry-pi-4-product-brief.pdf. [online] accessed on: September 2024.

\bibitem{smartmeter}
2016 {Engelec} {Power} {Technology} {Co.}, {Ltd}. {EEDDS238-4W} single phase {DIN} rail {WiFi} energy meter,\\ https://www.engelecgroup.com/eedds238-4w-single-phase-din-rail-wifi-energy-meter-pd49263996.html. [online] accessed on: September 2024.

\bibitem{loranode}
Hoperef {Electronic}. {RFM95/96/97/98(W)} - low power long range transceiver module v1.0,\\ https://www.mouser.be/datasheet/2/975/1463993415rfm95\_96\_97\_98w-1858106.pdf. [online] accessed on: September 2024.

\bibitem{rbp3}
Raspberry {Pi} {Ltd}. {Raspberry} {Pi} 3 model {B}, \\ https://www.raspberrypi.com/products/raspberry-pi-3-model-b. [online] accessed on: September 2024.

\bibitem{loragateway}
Dragino {Technology} {Co}., {Ltd}. {Open} source {LoRa} {WiFi} gateway {LG01},\\ https://www.dragino.com/downloads/downloads/datasheet/en/datashe\\ et\_lg01.pdf. [online] accessed on: September 2024.

\bibitem{wemosmini}
2021-2024, wemos.cc. {LOLIN} {D1} mini,\\ https://www.wemos.cc/en/latest/d1/d1\_mini.html. [online] accessed on: September 2024.

\bibitem{polar}
Polar. {Polar} {M600}, \\ https://support.polar.com/e\_manuals/m600/wear-os/polar-m600-user-manual-english/manual.pdf. [online] accessed on: September 2024.

\bibitem{imagesensor}
2024 {Omnivision}. ({End} of life) {CMOS} {VGA} (640x480) image sensor with {OmniPixel} 2 technology,\\ https://www.ovt.com/products/ov7725. [online] accessed on: September 2024.

\bibitem{nrf}
2024 {Nordic} {Semiconductor}. {nRF52840},\\ https://www.nordicsemi.com/products/nrf52840. [online] accessed on: September 2024.

\bibitem{jetson}
Nvidia. {Jetson} nano {2GB} developer kit, \\ https://developer.nvidia.com/embedded/jetson-nano-2gb-developer-kit. [online] accessed on: September 2024.

\bibitem{intel}
Intel. {Intel} core {i3-6006U} processor specification, \\ https://ark.intel.com/content/www/us/en/ark/products/91157/intel-corei3-6006u-processor-3m-cache-2-00-ghz.html. [online] accessed on: September 2024.

\bibitem{lenovo}
Lenovo. {Legion} 5i 17, \\ https://www.lenovo.com/in/en/p/laptops/legion-laptops/legion-5-series/legion-5i-17/88gmy501441?orgref=https\%253a\%252f\%252f. [online] accessed on: September 2024.

\bibitem{googlepixel}
Wikipedia. {Pixel} 2, \\ https://en.wikipedia.org/wiki/pixel\_2. [online] accessed on: September 2024.

\bibitem{lecun1998gradient}
Y.~LeCun, L.~Bottou, Y.~Bengio, P.~Haffner, Gradient-based learning applied to document recognition, Proceedings of the IEEE (1998).

\bibitem{kim2023online}
Y.~Kim, B.~Jung, Y.~Song, Online {Learning} for joint energy harvesting and information decoding optimization in {IoT}-enabled smart city, IEEE Internet of Things Journal (2023).

\bibitem{wang2023dual}
Y.~Wang, F.~Shang, J.~Lei, X.~Zhu, H.~Qin, J.~Wen, Dual-attention assisted {Deep} {Reinforcement} {Learning} algorithm for energy-efficient resource allocation in {Industrial} {Internet} of {Things}, Future Generation Computer Systems (2023).

\bibitem{wei2021intelligent}
K.~Wei, L.~Zhang, S.~Wang, Intelligent channel allocation for {Age} of {Information} optimization in {Internet} of {Medical} {Things}, Wireless Communications and Mobile Computing (2021).

\bibitem{wang2020big}
T.~Wang, H.~Ke, X.~Zheng, K.~Wang, A.~Sangaiah, A.~Liu, Big data cleaning based on mobile {Edge} {Computing} in industrial sensor-cloud, IEEE Transactions on Industrial Informatics (2020).

\bibitem{cao2020novel}
L.~Cao, Y.~Cai, Y.~Yue, S.~Cai, B.~Hang, A novel data fusion strategy based on {Extreme} {Learning} {Machine} optimized by bat algorithm for mobile heterogeneous {Wireless} {Sensor} {Networks}, IEEE Access (2020).

\bibitem{caba2018testing}
J.~Caba, F.~Rinc{\'o}n, J.~Dondo, J.~Barba, M.~Abaldea, J.~C. L{\'o}pez, Testing framework for in-hardware verification of the hardware modules generated using {HLS}, International Symposium on Power and Timing Modeling, Optimization and Simulation (2018).

\bibitem{gudnavar2023edge}
A.~Gudnavar, K.~Naregal, {Edge} {Computing} in {Internet} of {Things} ({IoT}): Enhancing {IoT} ecosystems through distributed intelligence, Advancement of IoT in Blockchain Technology and its Applications (2023).

\bibitem{donoho2024data}
D.~Donoho, Data science at the singularity, Harvard Data Science Review (2024).

\bibitem{gloria2017comparison}
A.~Gl{\'o}ria, F.~Cercas, N.~Souto, Comparison of communication protocols for low cost {Internet} of {Things} devices, South Eastern European Design Automation, Computer Engineering, Computer Networks and Social Media Conference (2017).

\bibitem{huang2006extreme}
G.-B. Huang, Q.-Y. Zhu, C.-K. Siew, Extreme {Learning} {Machine}: Theory and applications, Neurocomputing (2006).

\bibitem{lukovsevivcius2009reservoir}
M.~Luko{\v{s}}evi{\v{c}}ius, H.~Jaeger, Reservoir {Computing} approaches to {Recurrent} {Neural} {Network} training, Computer Science Review (2009).

\bibitem{tavanaei2019deep}
A.~Tavanaei, M.~Ghodrati, S.~R. Kheradpisheh, T.~Masquelier, A.~Maida, Deep {Learning} in {Spiking} {Neural} {Networks}, Neural Networks (2019).

\bibitem{young2019review}
A.~R. Young, M.~E. Dean, J.~S. Plank, G.~S. Rose, A review of spiking neuromorphic hardware communication systems, IEEE Access (2019).

\bibitem{matsubara2022split}
Y.~Matsubara, M.~Levorato, F.~Restuccia, Split {Computing} and {Early} {Exiting} for {Deep} {Learning} applications: Survey and research challenges, ACM Computing Surveys (2022).

\bibitem{6gwhitepaper}
B.~Rong, {6G}: The next horizon: From connected people and things to connected intelligence, IEEE Wireless Communications (2021).

\bibitem{biason2017eccentric}
A.~Biason, C.~Pielli, M.~Rossi, A.~Zanella, D.~Zordan, M.~Kelly, M.~Zorzi, {EC-CENTRIC}: An energy- and context-centric perspective on {IoT} systems and protocol design, IEEE Access (2017).

\bibitem{6gbricks}
F.~Minucci, R.~M. Noguera~Oishi, H.~Xiong, D.~Verbruggen, C.~Thys, R.~Hersyandika, R.~Beerten, A.~Colpaert, V.~Ranjbar, S.~Pollin, Building a real-time physical layer labeled data logging facility for {6G} research, IEEE International Workshop on Computer Aided Modeling and Design of Communication Links and Networks (2024).

\end{thebibliography}






\end{document}